\documentclass[twocolumn,preprintnumbers,prd,superscriptaddress,nofootinbib,floatfix,showpacs,showkeys]{revtex4-1}

\usepackage{float}
\usepackage{graphicx}  

\usepackage{multirow}
\usepackage{natbib}
\usepackage{pifont}
\usepackage{amssymb}
\usepackage{bm}        
\usepackage{amsfonts}  
\usepackage{amsmath}   
\usepackage{amssymb}   
\usepackage[titletoc]{appendix}
\usepackage{color}
\usepackage{hyperref}
\usepackage{cleveref}
\usepackage[english]{babel}
\usepackage{booktabs}
\usepackage{mathtools}
\usepackage{subfigure}
\usepackage{enumerate}

\definecolor{vividviolet}{rgb}{0.62, 0.0, 1.0}
\definecolor{amaranth}{rgb}{0.9, 0.17, 0.31}
\definecolor{palatinateblue}{rgb}{0.15, 0.23, 0.89}
\definecolor{brightpink}{rgb}{1.0, 0.0, 0.5}
\definecolor{cornflowerblue}{rgb}{0.39, 0.58, 0.93}
\definecolor{deepcarminepink}{rgb}{0.94, 0.19, 0.22}
\definecolor{radicalred}{rgb}{1.0, 0.21, 0.37}

\hypersetup{ linktoc=all,
    colorlinks, linkcolor={palatinateblue},
    citecolor={brightpink}, urlcolor={amaranth}
}

\begin{document}

\title{Comparing geometric and gravitational particle production in Jordan and Einstein frames}

\author{Alessio Belfiglio}
\email{alessio.belfiglio@unicam.it}
\affiliation{University of Camerino, Via Madonna delle Carceri, Camerino, 62032, Italy.}
\affiliation{Istituto Nazionale di Fisica Nucleare (INFN), Sezione di Perugia, Perugia, 06123, Italy.}

\author{Orlando Luongo}
\email{orlando.luongo@unicam.it}
\affiliation{University of Camerino, Via Madonna delle Carceri, Camerino, 62032, Italy.}
\affiliation{Istituto Nazionale di Fisica Nucleare (INFN), Sezione di Perugia, Perugia, 06123, Italy.}
\affiliation{Department of Mathematics and Physics, SUNY Polytechnic Institute, Utica, NY 13502, USA.}
\affiliation{INAF - Osservatorio Astronomico di Brera, Milano, Italy.}
\affiliation{Al-Farabi Kazakh National University, Almaty, 050040, Kazakhstan.}

\author{Tommaso Mengoni}
\email{tommaso.mengoni@unicam.it}
\affiliation{University of Camerino, Via Madonna delle Carceri, Camerino, 62032, Italy.}
\affiliation{Istituto Nazionale di Fisica Nucleare (INFN), Sezione di Perugia, Perugia, 06123, Italy.}
\affiliation{INAF - Osservatorio Astronomico di Brera, Milano, Italy.}

\begin{abstract} 
We explore cosmological particle production associated with inflationary fluctuations by comparing gravitational and geometric mechanisms, within a non-minimal Yukawa-like coupling between a inflaton and spacetime curvature. We show under which circumstances the number of geometric particles is comparable to the purely gravitational contribution by introducing the S-matrix formalism and approximating the inflationary potentials through power series. Despite the geometric production is a second-order effect, we show that it cannot be neglected \emph{a priori}, emphasizing that throughout inflation we do expect geometric particles to form as well as in the reheating phase. This result is investigated in both the Jordan and Einstein frames, selecting inflationary potentials supported by the Planck Satellite outcomes, \textit{i.e.}, the Starobinsky, hilltop, $\alpha$-attractor, and natural inflationary paradigms. We notice that, for large field approaches, the only feasible coupling may occur in case of extremely small field-curvature coupling constants, while the small field frameworks, namely the hilltop and $\alpha$-attractor models with quartic behavior, are essentially ruled out since they do not appear dynamically viable. We then point out that, while dynamically there exists a net equivalence between the Jordan and Einstein frames, the number of particles appears quite different passing from a representation to another one, thus pointing out the existence of a possible \emph{quantum frame issue}. In summary, while the classical inflationary dynamics is equivalently described in the Jordan and Einstein frames, a viable physical description of particle production valid in both the frames may be hindered by the absence of a complete quantum gravity. The latter could therefore be seen as missing puzzle to address the longstanding challenge of consistently passing between the two frames.
\end{abstract}

\pacs{03.70.+k,98.80.Cq, 98.80.-k, 98.80.Es}

\maketitle 
\tableofcontents

\section{Introduction}

Despite inflation remains the simplest paradigm to address the \emph{classical} Big Bang  issues   
\cite{riotto,Bassett_2006,Vazquez_Gonzalez_2020}, both the inflaton nature and the most viable inflationary potential are nowadays objects of heated debates  \cite{starob,higgs,hilltop,natural,universality,Linde_1994}.

Thus, numerous inflationary models have been proposed throughout the years \cite{Odintsov:2023weg,Bamba:2014daa, Berera:1995ie}, within the classes of  \emph{old} and \emph{new inflation} scenarios, respectively involving \emph{phase transitions} \cite{guth,linde,newinfl} and \emph{chaotic graceful exit} \cite{LINDE1983177}. Examples are \emph{thermodynamic inflation} \cite{Andrei,gtd},  multi-field inflation \cite{Senatore:2010wk}, spinor inflation \cite{Armendariz-Picon:2003wfx,spinor}, exotic approaches \cite{mengoni,D_Agostino_2022} and so on, see e.g. \cite{hybrid,vacuuminflation,DVALI199972}.

In a nutshell, single-field inflationary scenarios can be conventionally sorted out into \emph{small and large field} approaches. In the former case, inflation is caused by a scalar field gradually shifting from small values to larger ones, approaching the minimum of its potential \cite{hilltop,natural,natural1}. Conversely, for large field models, the process is inverted \cite{starob,1983SvAL....9..302S,ferrara}. 

In this respect, the most recent  Planck satellite findings \cite{planck} indicate that the best-fitting inflationary model is the large field Starobinsky potential \cite{starob}, turning out to be  \emph{conformally equivalent} to a non-minimally coupled field with Ricci curvature \cite{Kehagias:2013mya}.

Remarkably, one has not to marvel toward experimental  inconsistencies against chaotic  potentials, e.g.,  $V(\phi)=\frac{1}{4}\lambda\phi^4$ appears experimentally disfavored. However, before discounting this class of models, if a \emph{Yukawa-like} interacting Lagrangian between curvature and field is included,  a significant statistical enhancement occurs, \emph{naively suggesting that a possible non-minimal coupling could effectively take place at primordial times. }

Hence, exploring non-minimally coupled inflationary scenarios becomes essential to pin down the inflaton origin\footnote{Non-minimally coupled inflation gained attention about a decade ago, emerging within the framework of \emph{unified theories}. which are now questioning the validity of a few longstanding inflationary models, ruminating on how inflation can incorporate the dark sector, instead, see e.g. \cite{mengoni,D_Agostino_2022,Luongo_2018,Belfiglio:2023rxb,Belfiglio:2023eqi}.  }  \cite{improvement,Hertzberg_2010,density,complete,Karciauskas:2022jzd,obscons,preheating,futamase,futamase1,LUCCHIN1986163}.

Within this scenario, a perplexing task is understanding how particles can be created across the universe evolution. On the one hand, inflationary fluctuations lead to cosmological inhomogeneities during accelerated expansion. So, by virtue of Einstein's field equations, a pure geometric mechanism of particle production may occur, see Refs. \cite{friemanpart,cespedes}. On the other hand, field evolution during cosmic expansion produces particles directly from vacuum fluctuations, see Refs,  \cite{MUKHANOV1992203,Ford:2021syk,Parker,ema,superheavy,superheavy1}. This process, dubbed \emph{gravitational particle production}, is in general related to Bogoljubov transformations for quantum fields.

As a matter of fact, particles obtained from gravitational and geometric mechanisms are also associated with  entanglement entropy generation \cite{Ball:2005xa,entanglement,Martin-Martinez:2012chf,moradi}, whose properties could provide valuable insights into the expansion of the Universe \cite{Pierini:2018wki,Pierini:2015jma,Pierini:2016rkk,Luongo:2019ztg,Capozziello:2013wea,Luongo:2014qoa}.

For example, processes of geometric births of particles may lead to dark matter's origin at early times  \cite{Chung:2011ck,Ema:2019yrd,pion,gravitino}, either if induced by spectator fields \cite{spect,spectalpha} or, afterwards inflation, \textit{i.e.}, in the reheating  \cite{Belfiglio:2023eqi,superheavy,superheavy1}. Specifically, the interacting  Lagrangian may give rise to particles, dressed by the interaction, reviewed as \emph{geometric quasi-particles} \cite{Belfiglio:2022cnd,Belfiglio:2022yvs,Belfiglio:2023eqi,Belfiglio:2023rxb}.

In this paper, based on the above considerations, we compare geometric and gravitational particle productions within non-minimally coupled inflationary scenarios.
Thus, the first step is to introduce a Yukawa-like coupling between the inflaton field and scalar curvature for viable inflationary models, \textit{i.e.}, the Starobinsky\footnote{For the sake of completeness, the Starobinsky potential was originally introduced extending gravity through a Lagrangian density $\mathcal L\sim R+\alpha R^2$. Thus, adding further couplings to the Starobinsky potential may lead to conceptual problems. However, since during the years, several Starobinsky-like potentials have been obtained with different perspectives, it is possible to presume its origins from fields, rather than from extensions of general relativity, see e.g. Refs. \cite{D_Agostino_2022,mengoni}.
For this reason, we do not exclude \emph{a priori} the Starobinsky model non-minimally coupled with $R$. }, hilltop, $\alpha$-attractor potentials and the natural inflation \cite{1983SvAL....9..302S,starob,Mukhanov:1982nu,universality,ferrara1,hilltop,natural, natural1, Dolgov_1997}. For each potential, we then investigate whether non-minimal coupling interferes with inflationary dynamics and we definitely find that only the weakly interacting limit on the coupling constant turns out to be physically viable to describe inflation. This suggests that, although small-field models may be partially supported by experimental data, they appear disfavored to explain inflation. To better see this fact, we examine inflationary dynamics in both the Jordan and Einstein frames, assuming a perturbed de Sitter evolution during the slow-roll phase \cite{postma,Capozziello_1997}. Our intend is thus identifying the physical conditions for each potential allowing for equivalence between the two frames,  appearing quite different though. Afterwards, we study the dynamics of scalar inflaton fluctuations from an initial Bunch-Davies vacuum state \cite{Bunch:1978yq} at the onset of inflation up to the reheating stage, here formulated as a dust-like epoch\footnote{Including reheating is shown to  recover adiabaticity after the strongly non-adiabatic inflationary expansion and, so, the presence of an adiabatic expansion regime is in fact crucial to properly define a vacuum state, and consequently the notion of particle production, immediately after inflation \cite{Birrell:1982ix,Danielsson:2003wb,Brian}.}. Accordingly, we explore particle production associated with inflaton fluctuations in the above-depicted scenarios by investigating the Jordan and Einstein frames. Specifically, within a perturbative approach, we introduce a \emph{geometric particle production mechanism} during inflation. There, as the Bogoljubov coefficients are zero, through a numerical reconstruction of the involved inflationary potentials, we calculate the net  pairs and quartets exhibiting different momenta, behaving therefore differently with respect to the gravitational particle production scheme. These particles, moreover, are produced during inflation, instead of being evaluated at the end of its phase and, so, we remarkably show under which circumstances this novel approach to particle production might be taken into consideration significantly contributing alongside the more studied gravitational production. Last but not least, in view of the two particle production treatments, gravitational and geometric, we exploit their consequences to further investigate the frame equivalence problem. In other words, we compute particles in both the two frames and, knowing that the number of particles should be the same in principle, we show that the two frames cannot be fully-equivalent. Motivated by the fact that there exists a dynamical equivalence between the frames, albeit an evident non-equivalence in the number of produced particles, we propose strategies to fix this issue, underling the evident need for having a complete quantum gravity theory that would reconcile the particle production mechanisms in both the Jordan and Einstein frames. 

The paper is structured as follows: in Sect. \ref{Sect 2}, we introduce the non-minimally coupled inflationary dynamics, highlighting the difference between the Jordan and Einstein frames and showing results about the potentials considered. In Sect. \ref{sect 3}, we describe the evolution of inflaton fluctuations, while in Sect. \ref{Sect 4}, we introduce the particle production mechanisms. In Sect. \ref{Sect 5}, we present the outcomes of particle production for the models investigated, specifically comparing the two mechanisms and their results in both frames. Sect. \ref{sect 6} offers a detailed discussion of the main results and their physical implications. Finally, in Sect. \ref{sect 7}, we conclude with final outlooks and future perspectives.


\section{Non-minimal set up}\label{Sect 2}

There is no evidence \emph{a priori} suggesting that the inflaton might be minimally coupled to gravity, but rather indications in favor of non-minimal couplings arise in the contexts of Higgs inflation \cite{Garcia-Bellido:2008ycs,Barvinsky:2009fy}, quasiquintessence field \cite{mengoni} and so on. Consequently, there are several encouraging hints in favor of a non-minimal coupling between the inflaton and $R$, so that, when considering a single scalar field, we seek for the simplest interacting Yukawa-like term, yielding
\begin{align}
    \mathcal{L}&=\frac{1}{2}g^{\mu\nu}\partial_\mu\phi\partial_\nu\phi-V(\phi)+\frac{1}{2}\xi R\phi^2\,,\label{L tot}
\end{align}
where $1/2$ is arbitrary and the coupling constant, $\xi$, represents the strength, whose range spans from negative to positive values.
Shortly, the interacting term is hereafter denoted by $\mathcal{L}_{int}=\frac{1}{2}\xi R\phi^2$.

Accordingly, the whole interacting term can be recast in the form of an effective potential,  $V_{eff}(\phi)=V(\phi)-\frac{1}{2}\xi R\phi^2$, providing an effective gravitational constant, $G_{eff}$, coupled to $\phi$, as
\begin{equation}\label{geff}
G_{eff}=\frac{G}{1-\xi\chi\phi^2}\,,
\end{equation}
with $\chi\,\equiv\,8\pi/M_{Pl}^2=8\pi G$, valid within the inflationary scenario, assuming minimal gravity to be restored lately. 

Consistency with observations does not require gravity to behave differently at neither small nor large scales, implying the further condition, $|\xi|\ll1$. It is worth noting that the weak coupling case can be guaranteed  during inflation by just requiring the weaker condition $G_{eff}>0$, satisfying $\phi^2<\frac{1}{\chi\xi}$ \cite{preheating,futamase,Tsujikawa_2004}.

We will clarify later in the text (Sect. \ref{sect 3.1.2}) how the conformal and strong coupling regimes typically do not give rise to a well-defined inflationary stage.

\subsection{Singling out the frame}

The Yukawa-like coupling term introduced in Eq. \eqref{L tot} results in conceptual complications. Particularly, the main issue consists in choosing the frame, either Jordan or Einstein \cite{faraoni,Catena_2007,einjor}. 

Indeed, inflation can be described  in both the above frames \cite{postma,Capozziello_1997}, adopting the following features.
\begin{itemize}
    \item[-] In the Jordan frame, the action remains unaltered. The presence of the interacting term, $\mathcal L_{int}$, introduces a fifth force due to the non-minimal interaction.
    \item[-] In the Einstein frame, a conformal transformation is applied to the metric, eliminating the interaction.
\end{itemize}

Consequently, the ``frame issue" lies on wondering whether the fifth force exists or not. Accordingly, choosing one frame or another has potentially relevant implications, not fully understood nowadays \cite{faraoni2007}. 

Recently, it has been proved that the two frames yield equivalent results in describing the inflationary dynamics \cite{mengoni}. 

This can be seen as a consequence of the fact  that, for non-minimally coupled single scalar fields, quantum corrections in inflationary dynamics are negligible \cite{Hertzberg_2010}, suggesting to analyze  the dynamics through classical field equations.

Hence, it is natural to pursue the aim of confirming this result when extending the issue of choosing the right frame within the realm of particle production.


\subsection{The Jordan frame}\label{Sect 2.2}

Within this frame, the presence of non-minimal coupling modifies Einstein's field equations, as we explicitly show in Appendix \ref{appendix1}. 
In view of this modification, and bearing in mind that the background inflaton field is here  homogeneous, the first Friedmann equation approximately reads
\begin{equation}\label{fried jordan}
    H^2\,\simeq\,\frac{\chi}{3(1-\chi\xi\phi^2)}\left(\frac{1}{2}\dot\phi^2+V(\phi)\right)\,.
\end{equation}
As a consequence, the continuity equation becomes
\begin{equation}
    \ddot\phi+{3}H\dot\phi+{ V^\prime(\phi)-\xi R\phi}\simeq0,\label{Jordan dyn}
\end{equation}
where the slow-roll assumption allows us to consider $R^\prime=\frac{\dot R}{\dot\phi}\simeq0$. Specifically, in a quasi-de Sitter phase, the scalar curvature, $R$,  
\begin{align}
    R=\frac{\chi\,\big[(1-6\xi)\dot\phi^2-4V(\phi)-6\xi\phi\ddot\phi-18\xi H\phi\dot\phi\big]}{1-\chi\xi\phi^2},\label{scalar curvature}
\end{align}
can be considered approximately constant, since $\ddot\phi$ represents a second-order term with respect to the others. This scenario, in principle, could give rise to problems when aiming to smoothly exit inflation towards the minimum of the effective potential.

In particular, in the case of small field inflation with a positive coupling strength, the interacting Lagrangian exhibits an opposite trend compared to the potential, while, for large field inflation, the situation is reversed.

In this respect, we thus need to examine both negative and positive couplings for each potential. Additionally, we explore whether the effective potentials under consideration lead to well-defined inflationary stages without losing consistency with observational data.


\subsection{The Einstein frame}\label{Sect 2.1}

In the Einstein frame, we recover a minimally coupled action given by
\begin{equation}
    S_E=\int{d^4x{\sqrt{-g^E}}\left[-\frac{R^E}{2\chi}+\frac{1}{2}g^{\mu\nu}_{E}\partial_\mu h\partial_\nu h-V_E(h)\right]},
\end{equation}
where, as reported in Appendix \ref{appendix1},  $h$ is a new field introduced as a consequence of the conformal transformation, see Eq. \eqref{h}, whereas the potential turns out to be 
\begin{equation}
    V_E(h)=\frac{V(\phi(h))}{\left(1-\xi\chi\phi^2(h)\right)^2}.
\end{equation}
Specifically, the field transformation from $\phi$ to $h$ is, in general, not analytic. Three main cases can be thus explored,  
\begin{subequations}
    \begin{align}
    \phi&=\frac{1}{\sqrt{\chi\xi}}\tanh(\sqrt{\chi\xi}h), &\xi=1/6,\label{conformal coupling}\\
    \phi&\simeq\frac{1}{\sqrt{\chi\xi}}\sin(\sqrt{\chi\xi}h),&0<\xi\ll{1},\label{small xi}\\
    \phi&\simeq-\frac{1}{\sqrt{\chi\xi(6\xi-1)}}\sinh\left(\frac{\sqrt{\chi\xi}h}{\sqrt{6\xi-1}}\right),&-1\ll\xi<0.\label{small neg xi}
    \end{align}
\end{subequations}
where the first is known as conformal coupling and results in analytical solutions \cite{futamase1}, whereas the others are associated with weak interacting limiting cases.

\subsubsection{Inflationary dynamics in the Einstein frame}

We now focus on describing inflationary dynamics regardless the $\xi$ sign. 
Lying on the Einstein frame, the Hilbert-Einstein Lagrangian is recovered, and so the dynamics appears quite straightforward, providing the new scalar field, $h$, and the transformed potentials, $ V_E(h)$, only.

The field evolution can be recovered starting from the Lagrangian in Eq. \eqref{L tot}. The associated continuity equation reads
\begin{equation}\label{infl dyn}
    \ddot h+3H\dot h\,=\,-V_{E}^{\prime}(h)\,.
\end{equation}
This condition, along with the Friedmann equations, fully describes the inflaton dynamics.

Inflationary background evolution can be modeled in terms of a quasi-de Sitter expansion  \cite{Riotto:2010jd}, where the corresponding Friedmann and continuity equations become, respectively
\begin{subequations}
    \begin{align}
    H^2&\simeq\frac{8\pi}{3M_{Pl}^2}{V}_E(h),\label{srfr}\\
    3H&\dot\phi\simeq-V_{E}^{\prime}(h).\label{srdin}
\end{align}
\end{subequations}
The main features of slow-roll evolution are encoded in the slow-roll parameters
\begin{subequations}
\begin{align}
    \epsilon&\equiv-\frac{\dot H}{H^2}\simeq\frac{M_{Pl}^2}{16\pi }\left(\frac{ V_E^\prime}{ V_E}\right)^2,\label{eps}\\ 
    \eta&\equiv-\frac{\ddot h}{H\dot h} -\frac{\dot H}{H^2} \simeq\frac{M_{Pl}}{8\pi}\left(\frac{ V_E^{\prime\prime}}{ V_E}\right).\label{eta}
\end{align}
\end{subequations}
The latter expressions are the first-order approximated terms, usually dubbed potential slow-roll parameters \cite{Ellis_2015}. 

Specifically, the slow-roll phase, and inflation, occur as long as $\ddot{a}>0\>\iff\>\epsilon,\eta\ll1$. Nevertheless, accounting second-order terms in the approximation, it can be shown that inflation ends as 
\begin{equation}\label{infl end}
    \epsilon\simeq\left(1+\sqrt{1-\frac{\eta}{2}}\right)^2,
\end{equation}
that turns out to better approximate the inflationary stage. 

Inflation is typically characterized by the number of e-foldings that, within the slow-roll approximation, are easily computed. 

Specifically, denoting the values of the inflaton field at the beginning and end of inflation as $\phi_i$ and $\phi_f$,  respectively, the total number of e-foldings, at first-order, reads
\begin{equation}
    N\equiv\int_{\tau_i}^{\tau_f}{Hd\tau}\simeq-\frac{8\pi}{M_{Pl}^2}\int_{\phi_i}^{\phi_f}{\frac{ V_E}{ V_E^\prime}d\phi},\label{efold slowroll1}
\end{equation}
whereas at second-order approximation, we end up with 
\begin{equation}\label{efold slowroll}
    \Tilde{N}\simeq-\sqrt{\frac{4\pi}{M_{Pl}^2}}\int_{\phi_i}^{\phi_f}{\frac{1}{\sqrt{\epsilon(\phi)}}\left(1-\frac{1}{3}\epsilon(\phi)-\frac{1}{3}\eta(\phi)\right)d\phi},
\end{equation}
where, by definition, the condition $\Tilde N\leq N$ holds.

To have a sufficient amount of inflation, one typically requires $\Tilde N\gtrsim70$ \cite{linde2005particle,Liddle_1994}. 
So, for the sake of convenience, we substitute $N$ with $\tilde N$ and, so, as the slow-roll phase ends, we expect the inflaton field to oscillate around the minimum of its potential, denoted as $\phi_0$. This phenomenon, which should represent the starting point of the reheating phase \cite{Kofman:1997yn,Allahverdi_2010}, then requires a \emph{graceful exit} from inflation \cite{LINDE1983177} and thus an appropriate form of the inflationary potential, as shown by Eq. \eqref{infl dyn}.

From an experimental viewpoint, the Planck mission data \cite{planck} offer valuable insights into inflationary models, specifically through the analysis of cosmic microwave background anisotropies and their power spectrum. Two crucial parameters, the tensor-to-scalar ratio $r$ and the spectral index $n_s$ \cite{Lyth_1999, Cook_2015}, play a fundamental role in constraining the characteristics of inflationary models. Within the slow-roll regime, these parameters are expressed as:
\begin{subequations}
    \begin{align}
    r&=16\epsilon^*,\label{r}\\
    n_s&=1+2\eta^*-6\epsilon^*.\label{ns}
    \end{align}
\end{subequations}
Here, the labels $*$ in the slow-roll parameters, $\epsilon^*$ and $\eta^*$, specify that they are evaluated at the horizon crossing for a fixed pivot scale, namely when, in the slow-roll regime, such a physical wavelength of fluctuations reaches the Hubble horizon.  Specifically, we here refer to as the pivot scale, $k_*= 0.002\,$Mpc$^{-1}$, adopted by the Planck mission and, accordingly, the horizon crossing is expected to occur at the beginning of the slow-roll phase. In this respect, we will test these parameters between an e-folding number ranging from $70$, marking the start of the slow-roll phase, to approximately $40$.

These fluctuations are strictly linked to cosmic microwave background anisotropies. Consequently, the Planck mission has provided various constraints on these parameters, as detailed in Tab. \ref{tab:}, contingent on the chosen background model and the specific dataset used.

\begin{table}[h]
  \centering
  \begin{tabular}{c|c|c}
    \hline\hline
    Data Set & $r$ Bound & $n_s$ Bound \\
    \hline
    Planck TT,TE,EE&&\\+lowEB+lensing & $<0.11$ & $0.9659\pm0.0041$  \\\hline
    Planck TT,TE,EE&&\\+lowE+lensing+BK15 & $<0.061$ & $0.9651\pm0.0041$ \\\hline
    Planck TT,TE,EE&&\\+lowE+lensing+BK15+BAO& $<0.063$& $0.9668\pm0.0037$ \\
    \hline\hline
  \end{tabular}
  \caption{Bounds were derived using three distinct datasets, with the $\Lambda$CDM+$r$ benchmark employed as the background model. These constraints are provided by the Planck mission \cite{planck}. }
  \label{tab:}
\end{table}


\subsection{Non-minimally coupled extensions of inflationary models}\label{Sect 2.3}

We heree introduce some potentials supported by Planck measurements and analyze whether the presence of non-minimal coupling can lead to a consistent inflationary scenario. In detail, our aim lies on 
 
 \begin{itemize}
     \item[-] testing non-minimally  inflationary dynamics and comparing the so-obtained tensor-to-scalar ratio and the the spectral index with Planck results, 
     \item[-] verifying under which conditions inflation holds for small, large and natural inflationary models,  
     \item[-] checking if the dynamics can be equivalent passing from the Jordan to the Einstein frame, 
     \item[-] reconciling the dynamical problem, \textit{i.e.}, having a non-minimal coupling during inflation, with the frame issue, namely the physical equivalence between the Jordan and Einstein representations. 
 \end{itemize}
 
While the last point will be discussed later, we here first focus on identifying parameters describing the inflationary dynamics, including the slow-roll parameters, Eqs. \eqref{eta} and \eqref{eps}, the e-folding number, Eq. \eqref{efold slowroll}, initial and final conditions, and the evolution of the inflaton field in configuration space.
We concentrate on the limit of weak interactions, $|\xi|\ll1, $ addressing both the Einstein and the Jordan frames. 

Thus, we perform the transformation between the field $\phi$ and $h$ using Eqs. \eqref{small xi} -- \eqref{small neg xi}, recovering the potentials in the Einstein frame, Eq. \eqref{pot}:
\begin{align}
     V_E(h)&=\frac{V(\phi(h))}{\left(1-\xi\chi\phi^2(h)\right)^2},\\
     \phi^{(+)}&\simeq\frac{1}{\sqrt{\chi\xi}}\sin(\sqrt{\chi\xi}h),\\
    \phi^{(-)}&\simeq-\frac{1}{\sqrt{\chi\xi(6\xi-1)}}\sinh\left(\frac{\sqrt{\chi\xi}h}{\sqrt{6\xi-1}}\right),
\end{align}
where the superscripts,  $(+)$ and $(-)$, specify the coupling constant sign.

To include small, large and natural inflationary paradigms, we here explore those  classes of potentials that fulfill the last findings by the Planck Satellite, as listed below.

\begin{itemize}
    \item[-] Among large field potentials, we select the Starobinsky model \cite{1983SvAL....9..302S,starob,Mukhanov:1982nu}, discussed in the introduction, and the $\alpha$-attractors models \cite{universality,ferrara1}, originally introduced within a supergravity context:
\begin{subequations} \label{larpot}
    \begin{align}
     V_{St}(\phi)&=\Lambda^4\left(1-e^{-\sqrt{\frac{2}{3}}\frac{\phi}{M_{Pl}}}\right)^2\,,\label{star}\\
     V_E^{(n)}(\phi)&=\Lambda^4\left[1-exp\left(-\sqrt{\frac{2}{3\alpha_n^E}}\frac{\phi}{M_{Pl}}\right)\right]^{2n}\,,\label{E pot}\\
     V_T^{(m)}(\phi)&=\Lambda^4\tanh^{2m}\left(\sqrt{\frac{1}{6\alpha_m^T}}\frac{\phi}{M_{Pl}}\right)\,.\label{T pot}
     \end{align}
\end{subequations}
Within the $\alpha$-attractors potentials, we limit our attention to the $E-$ and $T-$models, respectively indicated by the subscripts in the above equations, where $n,\,m=1,2$.
\item[-]  Conversely, concerning the class of small field, we examine the hilltop potentials \cite{hilltop}, introduced in the \emph{new inflation} paradigm. These potentials exhibit a slow-roll phase near the maximum of the potential and are given by
\begin{equation}
    V_{H}^{(p)}(\phi)\simeq\Lambda^4\left[1-\left(\frac{\phi}{\mu}\right)^p\right]\,,\label{hill pot}
\end{equation}
with $p=2;4$ corresponding to the quadratic and quartic hilltop inflation, respectively.
\item[-] Lastly, we investigate natural inflation \cite{natural, natural1, Dolgov_1997}, that can be traced back to a pseudo-Nambu Goldstone Boson undergoing a symmetry breaking mechanism through the following potential
\begin{equation}
    V_N(\phi)=\Lambda^4\left(1+\cos\left(\frac{\phi}{f}\right)\right)\,.\label{nat pot}
\end{equation}

\end{itemize}

For consistency, we set the free parameters as

\begin{subequations}
    \begin{align}
&\Lambda=10^{16}\,GeV\\ 
&\xi=\pm10^{-4},        
    \end{align}
\end{subequations}
whereas the free parameters of the models are set, according to Planck observations \cite{planck} as, 
\begin{subequations}
    \begin{align}
&\alpha_n^E=10,\\ 
&\alpha_n^E=0.1,\\ 
&f=1.5M_{Pl},
    \end{align}
\end{subequations}
while $\mu$ is not fixed \emph{a priori}.

Hence, 
\begin{itemize}
\item[-] the transformed potentials are displayed in Fig. \ref{fig pot},
\item[-] in the Einstein frame, we compute the slow-roll parameters using Eqs. \eqref{eps} -- \eqref{eta}, as shown in Fig. \ref{fig pot}, 
\item[-] consequently, we identify the slow-roll phase and determine the final value of the inflaton field, satisfying Eq. \eqref{infl end},
\item[-] we then obtain the initial condition for the field $h$, considering  $N\gtrsim70$.
\end{itemize}

Subsequently, from Eq. \eqref{srdin}, we are able to compute the initial condition on the time field variation,
\begin{equation}
    \dot h_{in}\simeq-\frac{ V^\prime(h_{in})}{3H_{in}}.
\end{equation}

In the Jordan frame, due to the presence of the Ricci scalar, the slow-roll parameters are no longer defined through Eqs. \eqref{eps} -- \eqref{eta}. Hence, we start from the previous initial conditions on the field $h$, from Eqs. \eqref{small xi} -- \eqref{small neg xi}, and  transform them into the Jordan frame to see whether they lead to consistent results
\begin{align}
    \dot\phi&\simeq\cos{(\sqrt{\chi\xi}h)}\dot h,\quad&\xi>0,\\
    \dot\phi&\simeq(1-6\xi)\cosh\left(-\frac{\sqrt{\chi\xi}h}{\sqrt{6\xi-1}}\right)\dot{h},\quad&\xi<0.
\end{align} 
All the initial and final conditions are reported in Tab. \ref{tab nonmin}.

\begin{table}[ht]
  \centering
  \begin{tabular}{c|c|c|c|c}
    \hline\hline
    {\bf$\xi^{(+)}$}&$h_{end}/M_{Pl}$ & $h_{in}/M_{Pl}$ & $\dot h_{in}/M_{Pl}^2\cdot 10^{-8}$& $H_I\,[GeV]\cdot10^{13}$\\
    \hline
    \hline
    $V_S$ &  $0.19$ & $2.36$ & $-3.01$& $2.03$  \\
    
    $V_E$ & $0.20$ & $2.99$ & $-4.14$ & $1.29$  \\
    
    $V_T$ & $0.19$ & $1.71$ & $-1.61$ & $2.37$  \\
    
    $V_N$ & $4.55$ & $1.74$ & $\,\,\,\,5.73$ & $2.81$ \\  
     \hline\hline
    {\bf$\xi^{(-)}$}&$h_{end}/M_{Pl}$ & $h_{in}/M_{Pl}$ & $\dot h_{in}/M_{Pl}^2$& $H_I\,[GeV]$\\
    \hline
    \hline
    $V_S$ &  $0.19$ & $2.30$ & $-2.62$ & $2.01$  \\
    
    $V_E$ & $0.20$ & $2.92$ & $-2.59$ & $1.25$  \\
    
    $V_T$ & $0.19$ & $1.66$ & $-1.40$ & $2.36$  \\
    
    $V_N$ & $4.46$ & $1.61$ & $\,\,\,\,5.76$ & $2.79$ \\ 
    \hline\hline
  \end{tabular}
  \caption{Resuming table with the initial and final conditions for the inflaton field in the Einstein frame. In the last column, the Hubble parameter during inflation, $H_I$, is reported. We set $\xi^{(+)}=-\xi^{(-)}=10^{-4}$.}  
  \label{tab nonmin}
\end{table}

In conclusion, we numerically address the continuity equations, Eqs. \eqref{Jordan dyn} -- \eqref{infl dyn}, respectively within the Jordan and Einstein frames, as illustrated in Fig. \ref{fig dyn} and therefore we can conclude as below. 

\begin{itemize}
    \item[-] The non-minimally coupled large field models manage to provide a suitable inflationary stage. The evolution of the inflaton field in the configuration space exhibits a slow-roll period, followed by well-defined attractor behavior around the minimum of the potential. During this phase, the field naturally converges toward the minimum, displaying a chaotic evolution.
    \item[-] Small fields appear quite problematic. Despite a Yukawa-like non-minimal coupling term can be accommodated in the majority of the above-depicted models, we observe as well that the hilltop potentials do not lead to any graceful exit, and that, similarly, $\alpha-$attractor model with $n=m=2$ present problems at the same stage, see Fig. \ref{fig not work}. Thus, we argue that non-minimal coupling does not fit well within small field potentials, especially when dealing with hilltop-like functional forms. Additionally, the fact that quartic exponential potentials do not suitably work provides further confirmation for Starobinsky-like inflationary potentials. 
    \item[-] Further, we compute the values of $r$ and $n_s$ of such extended inflationary models from the onset of inflation, \textit{i.e.}, $N\simeq70$, to $N\simeq40$. Hence, to assess the consistency of our study, we compare these values with Planck results. The plots of the tensor-to-scalar ratio and the spectral index reported in Fig. \ref{fig rns} satisfy the observational constraints. However, it is noteworthy that some models conform to specific dataset bounds rather than to all of them.
    \item[-] Finally, the analyses carried out in both the Einstein frame and the Jordan frame consistently validate the equivalence between the two frames. In other words, the inflationary dynamics and the corresponding predictions on observables \emph{are dynamically invariant with respect to the choice of frame}.
\end{itemize}

\subsubsection{The conformal coupling limit}\label{sect 3.1.2}

As previously mentioned, we also aim to demonstrate that, in the presence of strong field-curvature coupling, inflation is no longer possible. Considering the field transformation in Eq. \eqref{h}, we find that as the coupling strength $\xi$ increases, the dominant term becomes $\tanh{\left(\sqrt{\chi\xi}h\right)}$. In particular, in the limit of conformal coupling, \textit{i.e.}, when $\xi=1/6$, it is possible to recover the only possible analytic solution for the transformation, namely Eq. \eqref{conformal coupling}. Thus, we want to show that the case $|\xi|\simeq 1/6$ can be considered as an upper cutoff scale for the coupling constant.

We start by examining the presence of conformal coupling within the Starobinsky model, Eq. \eqref{star}.
Performing the transformation in Eq. \eqref{conformal coupling}, we obtain the new expression for the potential in the Einstein frame, Eq. \eqref{pot},
\begin{equation}
    V_{St}^{E}(\phi)=\Lambda^4\left(1-e^{-\sqrt{\frac{2}{3\chi\xi}}\frac{\tanh{\left(\sqrt{\chi\xi}h\right)}}{M_{Pl}}}\right)^2\cosh^4{\left(\sqrt{\chi\xi}h\right)}.
\end{equation}
First, we compute the slow-roll parameters, Eqs. \eqref{eps}, \eqref{eta}, asking whether they allow an inflationary stage.

\begin{figure}[ht]
  \centering\vspace{5mm}
\includegraphics[width=.95\linewidth]{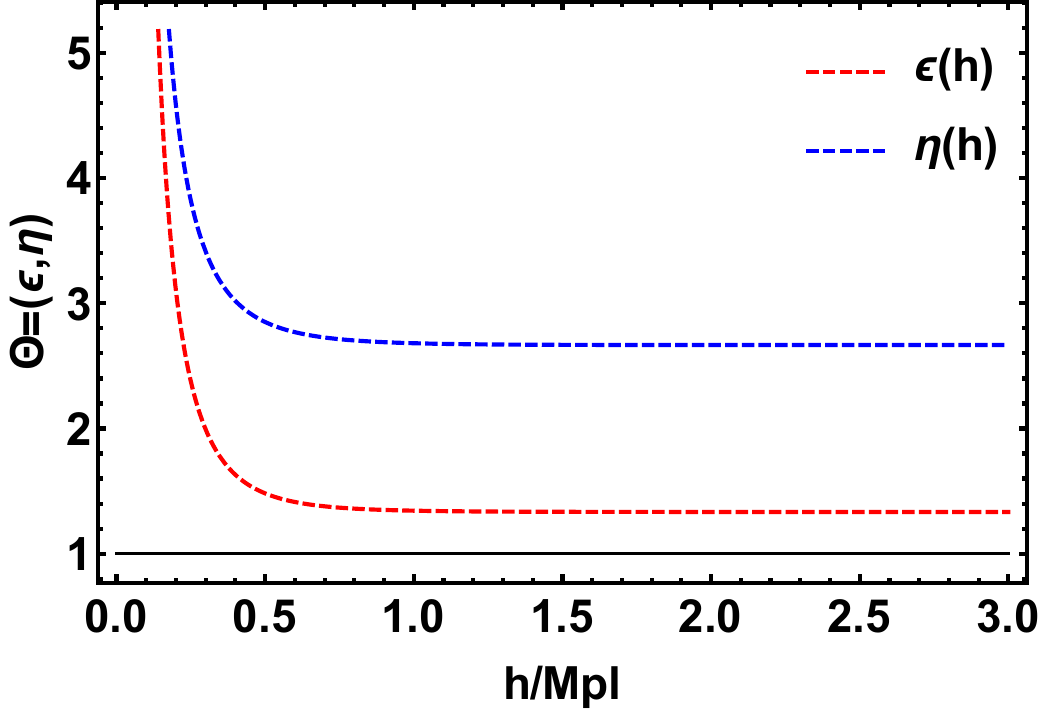}
\caption{The slow-roll parameters $\epsilon$ and $\eta$ of the Starobinsky potential conformally coupled with the scalar curvature $R$.}
\label{fig conf_coupl}
\end{figure}

As can be seen in Fig. \ref{fig conf_coupl}, their values never satisfies the condition $\epsilon,|\eta|\ll1$, thus indicating that inflation can not get started. We highlight that similar results can also be recovered for the other above-presented potentials. Thus, we conclude that conformal coupling is not a suitable choice. Additionally, larger values with respect to $|\xi|\simeq 1/6$ would result in slow-roll parameters that appear to be inconsistent with the latest  observational constraints.


\section{Dynamics of inflationary fluctuations}\label{sect 3}

Before discussing cosmological particle production originating from the aforementioned potentials, we need to introduce in the above-depicted scenario the quantum fluctuations inevitably associated with the presence of a scalar inflaton field. As we will see, such fluctuations are responsible for the geometric particle production mechanism that occurs \emph{during the inflationary epoch}. Specifically, we describe the inflaton fluctuations from the beginning of inflation to the end of reheating, when such fluctuations, along with the inflaton itself, are supposed to finish. Hence, the reheating phase is also included in order to compute the net geometric contribution to the number of particles. 
To better explain our treatment, we begin by characterizing the cosmological evolution and particle production in the Jordan frame. Afterwards, we explore how to recover the overall mechanism in the Einstein frame, with the intend of showing whether the two frames are equivalent or not.


\subsection{Perturbing the inflaton field}\label{sect 3.1}

We consider the usual ansatz where the inflaton field is split into a background homogeneous component, $\phi_0$, and a quantum fluctuation term, $\delta\phi$, \cite{Brandenberger:2003vk,MUKHANOV1992203}
\begin{equation}
    \phi({\bf x},t)=\phi_0(t)+\delta\phi({\bf x},t). \label{field} 
\end{equation}
These fluctuations are responsible for particle production, as we will see. 
To simplify our analysis, we shift from cosmic time $t$ to conformal time $\tau$, performing the transformation,     $dt\,=\,a(\tau)d\tau$  \cite{baumann2012tasi}. 

To handle scalar perturbations to linear order, we begin with the Friedmann-Robertson-Walker metric in conformal time and then adopt the widely-employed \emph{longitudinal}, or \emph{conformal Newtonian}, gauge \cite{Boya}, which gives
\begin{equation}\label{perturbed metric}
    g_{\mu\nu}=g_{\mu\nu}^{(0)}+\delta g_{\mu\nu}=a(\tau)^2\left(\eta_{\mu\nu}+2{ \Psi}\delta_{\mu\nu}\right),
\end{equation}
where $ \Psi$ represents the perturbation potential, fulfilling $|{\Psi}| \ll 1$ and thus ensuring the validity of our perturbative approach.

The inflaton equation of motion, corresponding to the Klein-Gordon equation of a scalar field in curved spacetime, is given by \cite{Duncan:1977fc}
\begin{equation}\label{kg}
    \frac{1}{\sqrt{-g}}\partial_\mu\left(\sqrt{-g}g^{\mu\nu}\partial_{\nu}\phi\right)=-\frac{\partial V_{eff}}{\partial\phi},
\end{equation}
which, via Eq. \eqref{field}, readily gives the background equation
\begin{equation}\label{fieldback}
\phi^{\prime\prime}_0+2\mathcal{H}\phi^\prime_0-\nabla^2\phi_0\,=\,-a^2V_{eff}^\prime.
\end{equation}
We can then expand the field fluctuations $\delta\phi$ and the perturbation potential $\Psi$ in Fourier modes as 
\begin{subequations}
    \begin{align}
    \delta\phi&=\frac{1}{\sqrt{V}}\sum_k\delta\phi_k(\tau)e^{i{\bf k\cdot x}},\\
    \Psi&=\frac{1}{\sqrt{V}}\sum_k\Psi_k(\tau)e^{i{\bf k\cdot x}}.
\end{align}
\end{subequations}
and, further, quantize the inflaton fluctuations by
\begin{equation}\label{fourierquant}
    \hat{\delta\phi}=\frac{1}{(2\pi)^{3/2}}\int{d^3k\left(\hat{a}_k\delta\phi_k(\tau)e^{i{\bf k\cdot x}}+\hat{a}_k^\dagger\delta\phi_k^*(\tau)e^{-i{\bf k\cdot x}}\right)},
\end{equation}
where $\hat{a}_k$ satisfies the canonical commutation relations $[\hat{a}_k,\hat{a}_{k^\prime}^\dagger]=\delta^{(3)}({\bf k}-{\bf k^\prime})$.

After some calculations, and denoting $\phi_0$ as $\phi$ from now, for simplicity, the equation of motion for the field fluctuations at first order can be obtained from Eq. \eqref{kg}, giving
\begin{align} \label{fluctmotion}
&\delta \phi_k^{\prime \prime}+ 2 \mathcal{H} \delta \phi_k^\prime+ k^2 \delta \phi_k-4 \Psi_k^\prime \phi^\prime \notag \\
&\simeq -\xi \left( -2k^2 \Psi_k-6\Psi_k^{\prime \prime}-24 \mathcal{H} \Psi_k^\prime-12 \frac{a^{\prime \prime}}{a} \Psi_k  \right) \phi \notag \\
&\ \ \ - \left( V^{\rm eff}_{, \phi \phi} \delta \phi_k + 2 \Psi_k V^{\rm eff}_{, \phi}  \right) a^2,
\end{align}
where we set $\frac{\partial}{\partial\tau}\equiv{}^\prime$, while $\mathcal{H}$ is defined as the Hubble parameter in conformal time, namely $\mathcal{H}\equiv a^\prime/a$ \cite{riotto}.
Assuming a weakly interacting regime, $|\xi|\ll1$, the perturbation potential satisfies
\begin{subequations}
    \begin{align}
    &\Psi_k^\prime+\mathcal{H} \Psi_k\,\simeq\, \epsilon \mathcal{H}^2 \frac{\delta \phi_k}{\phi^\prime}, \label{psipsi0}\\
    &\Psi_k^{\prime \prime}+2\left(\mathcal{H}-\frac{\phi^{\prime \prime}}{\phi^\prime} \right) \Psi_k^\prime+2\left( \mathcal{H}^\prime-\mathcal{H} \frac{\phi^{\prime \prime}}{\phi^\prime}\right)\Psi_k +k^2 \Psi_k\,\simeq\,0\,,\label{psipsi}
    \end{align}
\end{subequations}
at  first order  approximation.

During a slow-roll phase with $\epsilon\ll1$,  using Eq. \eqref{fieldback}, we can safely rewrite Eq. \eqref{fluctmotion} as
\begin{equation}\label{fluctmotion1}
    \delta \phi_k^{\prime \prime}+2 \mathcal{H} \delta \phi_k^\prime+\left[ k^2 + V^{\rm eff}_{, \phi \phi}a^2 +6 \epsilon \mathcal{H}^2 \right] \delta \phi_k = 0.
\end{equation}

\subsection{Evolution of fluctuations}

To explicitly derive the Fourier modes of inflaton fluctuations, we select a \emph{quasi-de Sitter} evolution \cite{Singh:2013dia}
\begin{equation}\label{desitter}
    a(\tau)= -\frac{1}{H_I (\tau-2\tau_f)^{1+\epsilon}},
\end{equation}
where $\tau_f$ represents the final time of the inflationary stage, while the slow-roll parameter $\epsilon$, according to the previous results, is taken constant and equal to $\epsilon=0.003$. Moreover, $H_I$ is the approximate constant value of the Hubble parameter during inflation. We underline that, within a single scalar field inflation, the ideal de Sitter spacetime description is only approximately valid during the slow-roll phase, thus we consider a quasi-de Sitter expansion to be more consistent with the inflationary framework. Nevertheless, particles are not supposed to be created as the universe undergoes a genuine de Sitter phase. In any cases, the parameter $H_I$, \textit{i.e.}, dominating the universe dynamics through the strong speed up,  can be computed from the slow-roll approximation and, in particular, the values corresponding to the underlying potentials are shown in Tab. \ref{tab nonmin}.

Noting that  $R=-6a^{\prime\prime}/a^3\simeq-6(2+3\epsilon)/(\tau a)^2$ and rescaling the field by $\delta\chi_k=a(\tau)\delta\phi_k$, Eq. \eqref{fluctmotion1} gives
\begin{equation}
    \delta \chi_k^{\prime \prime}+ \left[ k^2-\frac{1}{\tau^2} \left( \left( 1-6\xi\right)(2+3\epsilon)+6\epsilon- \frac{ V_{\phi \phi}}{H_I^2} \right)    \right] \delta \chi_k=0, \label{jor dyn pp}
\end{equation}
where, in agreement with the dynamical analyses of Sect. \ref{Sect 2}, we exploited $|R|^\prime\ll1$. 
Defining the parameter $\nu$ as 
\begin{equation}\label{nu}
    \nu\equiv \sqrt{\frac{1}{4}+(1-6\xi)(2+3\epsilon)+6\epsilon-V_{,\phi \phi}/H_I^2},
\end{equation}
we can finally express the field fluctuation modes as
\begin{equation}\label{generic sol}
    \delta \chi_k(\tau)= \sqrt{-\tau} \left[ C_1(k) H_{\nu}^{(1)}(-k\tau)+C_2(k) H_{\nu}^{(2)}(-k\tau)  \right],
\end{equation}
where $H_\nu^{(1)}$ and $H_\nu^{(2)}$ are the Hankel functions.

\subsection{Origin of fluctuations}

In order to determine the integration constants $C_1(k)$ and $C_2(k)$, we select the Bunch-Davies vacuum state as a proper boundary condition at the beginning of inflation, thus setting
\begin{equation}\label{bd}
    \lim_{\tau \to -\infty}\delta \chi_k(\tau)=\frac{e^{-ik\tau}}{\sqrt{2k}}.
\end{equation}
It must be noted that the problem of properly defining vacuum (and particle) states represents a relevant issue of quantum field theory in curved spacetime, due to the breaking of Poincaré symmetry \cite{Ford:2021syk}. In particular, since in inflationary cosmology all pre-existing fluctuations are red-shifted by spacetime acceleration, one typically assumes that the inflaton field starts mode by mode in its vacuum state \cite{Brandenberger:2003vk}.

Notably, it has been demonstrated that the Bunch-Davies vacuum state consists in a local attractor in the space
of initial states for expanding background \cite{enden} and it is then widely employed in inflationary scenarios.

Furthermore, we assume that all the modes of interest are within the cosmological horizon as inflation starts. From a Bunch-Davies initial state, they evolve until their wavelength reaches and exceeds the Hubble scale, occurring as $k > a(\tau) H$. At this point, they freeze out and escape the horizon, so these fluctuations are called super-Hubble modes, and they re-enter the horizon during later stages of the universe's evolution, directly contributing to the formation of large-scale structures. Modes, that do not cross the horizon, \textit{i.e.}, when  $k < a(\tau)H$, are known as sub-Hubble modes and remain in causal connection with the universe \cite{riotto,belfiglio2023superhorizon}. This distinction, in general, provides valuable insights across various topics, and in this work, we explore its impact on particle production.

Coming back to the vacuum state, Eq. \eqref{bd}, and matching it with the field expression, Eq. \eqref{generic sol}, we obtain the expression for the field fluctuations
\begin{equation}\label{gen sol field}
    \delta \phi_k (\tau)= \frac{\sqrt{-\pi\tau}}{2} e^{i\left( \nu+ \frac{1}{2}\right) \frac{\pi}{2}}  H^{\left(1\right)}_{\nu}\left(-k\tau\right) /a(\tau).
\end{equation}
To show it explicitly, we consider the limits of the Hankel functions, and consequently the behavior of the fluctuations modes
\begin{widetext}
\begin{align}
&\delta \phi_k^{\rm sub} \simeq \frac{1}{\sqrt{2k}}  \frac{e^{i \left( \nu+ \frac{1}{2} \right) \frac{\pi}{2}}\ e^{i\left( -k\tau-\frac{\pi}{2}\nu-\frac{\pi}{4}  \right)}}{a(\tau)},\quad&\quad
&H^{\left(1\right)}_{\nu}\left(x\gg1\right) \simeq\sqrt{\frac{2}{\pi x}}e^{i\left(x-\frac{\pi}{2}\nu-\frac{\pi}{4}\right)},\label{subfluc}
\\
&\delta \phi_k^{\rm super} \simeq e^{i\left(\nu-\frac{1}{2}\right)\frac{\pi}{2}}2^{\left(\nu-\frac{3}{2}\right)}\frac{\Gamma\left(\nu\right)}{\Gamma\left(\frac{3}{2}\right)}\frac{H_{I}}{\sqrt{2k^{3}}}\left(\frac{k}{aH_{I}}\right)^{\frac{3}{2}-\nu},\quad&\quad
&H^{\left(1\right)}_{\nu}\left(x\ll1\right)\simeq \sqrt{\frac{2}{\pi}}e^{-i\frac{\pi}{2}}2^{\left(\nu-\frac{3}{2}\right)}\frac{\Gamma\left(\nu\right)}{\Gamma\left(\frac{3}{2}\right)}x^{-\nu}.\label{supfluc}
\end{align}
\end{widetext}

As we will see later, despite the general agreement on the Bunch-Davies vacuum, it presents some critical issues. In this regard, the literature includes approaches where the initial conditions are chosen differently, see e.g.  \cite{Ashoorioon_2014,goldstein,PhysRevD.110.103502}, where initial states quite different from the Bunch-Davies one are adopted.
 

\subsection{Adiabatic expansion in the reheating era}

Before dealing with particle production mechanisms, we briefly describe the reheating phase, subsequent to inflation \cite{Kofman:1997yn}. During this stage the inflaton couples to other quantum fields and gradually vanishes, through a phase of adiabatic evolution.

Reheating is then characterized by oscillations of the inflaton around the minimum of the potentials addressed. Thus, by considering the average values of the quantities involved, it follows that the average of the equation of state $\langle\omega\rangle$ and pressure, $\langle P\rangle$ behave dust-like, namely  $\langle\omega\rangle\propto\langle P\rangle\simeq 0$.
Generally speaking, reheating is an intermediate era between inflation and radiative era, so we expect $\omega_{re}\in (-1/3,1/3)$ \cite{Cai:2015soa}.
If $\omega_{re}=1/3$, we would already be in radiation domination, meaning that reheating would be instantaneous, \textit{i.e.}, $N_{re}=0$.
Conversely, as shown in Ref. \cite{Ueno:2016dim}, we expect reheating  not to exceed $N_{re}\simeq30$.
Thus, we model the reheating phase on average as a matter-dominated stage, described by
\begin{equation}\label{reheating}
    H(t)\simeq H_{re}a^{-3/2}_{re}(t),\,\Rightarrow a(\tau)\simeq\frac{H_{re}^2}{4}\tau^2,
\end{equation}
where $H_{re}$ might be determined. 

Recalling that the universe subsequently enters a radiation-dominated epoch, which is well constrained by observations, we characterize this phase by
\begin{equation}\label{radiation}
    H(t)\simeq H_{rad}a^{-2}_{rad}(t),\,\Rightarrow a(\tau)\simeq H_{rad}\tau,
\end{equation}
with $H_{rad}=10^{-44}$ GeV \cite{Boya}.

Hence, by imposing continuity between these epochs, we obtain 
\begin{equation}
    \begin{cases}
    {\lim}_{\tau \to \tau_f^-} \, H(\tau)\,=\,{\lim}_{\tau \to \tau_f^+} \, H(\tau), \\\vspace{2mm}
    {\lim}_{\tau \to \tau_r^-} \, H(\tau)\,=\,{\lim}_{\tau \to \tau_r^+} \, H(\tau), \\\vspace{2mm}
    0<N_{re}<30,
\end{cases}
\end{equation}
where inflation ends at $\tau_f$, and where $\tau_r$ marks the transition from reheating to the radiation-dominated era. From this, it is straightforward to derive 
\begin{align}
    &\tau_f<\tau_r\leq \tau_fe^{15},\label{range of reheating}\\
    &\tau_f\in\left(\frac{1}{\sqrt{H_{rad}H_I}},\,\frac{e^{{15\over2}}}{\sqrt{H_{rad}H_I}}\right),\label{time rehe}
\end{align}
where we exclude the boundary values, because they correspond to the aforementioned limiting cases.

To ensure a well-defined \emph{out}-vacuum state and a proper definition of particles after inflation, an adiabatic regime must be recovered \cite{Birrell:1982ix,Parker}.
Accordingly, if the universe undergoes adiabatic expansion before the beginning of the radiative stage, where the fluctuations of the inflaton field and the field itself vanish, we could properly study particle production associated with inflaton fluctuations up to $\tau=\tau_r$. 
To check this, we restart from the equation of motion for the fluctuation modes, Eq. \eqref{fluctmotion1}, in terms of $\delta\chi_k(\tau)$ becomes
\begin{equation}\label{ll}
    \delta \chi_k^{\prime \prime}+\left[ k^2 + V_{, \phi \phi}a^2 -6 \epsilon \mathcal{H}^2-(1-6\xi)\frac{a^{\prime\prime}}{a} \right] \delta \chi_k = 0.
\end{equation}
Then, to account for the reheating scale factor, we substitute Eq. \eqref{reheating} into \eqref{ll}, yielding 
\begin{equation}\label{ll1}
    \delta \chi_k^{\prime \prime}+\left[ k^2 + V_{, \phi \phi}\frac{H_{re}^4}{16}\tau^4 -\frac{2}{\tau^2}(1-6\xi+12\epsilon) \right] \delta \chi_k = 0.
\end{equation}
Defining 
\begin{equation}
    \omega_k^2(\tau)\equiv  k^2 + V_{, \phi \phi}\frac{H_{re}^4}{16}\tau^4 -\frac{2}{\tau^2}(1-6\xi+12\epsilon),
\end{equation}
we expect field fluctuations to follow Eq. \eqref{ll1}, with the corresponding asymptotic behavior \cite{Boya, Rai:2020edx}
\begin{equation}\label{adiab eq}
    f_k(\tau)=\frac{e^{-i\int_{\tau_f}^\tau W_k(\tau^\prime)d\tau^\prime}}{\sqrt{2W_k(\tau)}}\,\rightarrow\frac{e^{-i\int_{\tau_f}^\tau \omega_k(\tau^\prime)d\tau^\prime}}{\sqrt{2\omega_k(\tau)}}.
\end{equation}
 This solution holds for $W_k^2(\tau)$ defined as
\begin{equation}
    W_k^2(\tau)=\omega^2_k(\tau)-\frac{1}{2}\left[\frac{W_k^{\prime\prime}(\tau)}{W_k(\tau)}-\frac{3}{2}\left(\frac{W_k^\prime(\tau)}{W_k(\tau)}\right)^2\right],
\end{equation}
whose expansion, up to second order, is given by
\begin{equation}
    W_k^2(\tau)=\omega^2_k(\tau)\left[1-\frac{1}{2}\frac{\omega_k^{\prime\prime}(\tau)}{\omega_k^3(\tau)}+\frac{3}{4}\left(\frac{\omega_k^\prime(\tau)}{\omega^2_k(\tau)}\right)^2\right].
\end{equation}
It can be now shown that, for super-Hubble modes, adiabaticity is recovered  if the following condition holds \cite{Birrell:1982ix,Rai:2020edx} 
\begin{equation}\label{adiab}
    \frac{\omega_k^\prime(\tau)}{\omega_k^2(\tau)}\simeq\frac{V_{, \phi \phi}\frac{H_{re}^4}{4}\tau^3}{V_{, \phi \phi}^2\frac{H_{re}^8}{256}\tau^8}=\frac{64}{V_{, \phi \phi}H_{re}^4\tau^5}\ll1,
\end{equation}
where we neglect all the second-order contributions.

For $\tau$ satisfying the condition in Eq. \eqref{adiab}, we can properly define particles in the \emph{out}-region, as adiabaticity is finally achieved.

Comparing this result with the range of times obtained for the reheating stage, Eq. \eqref{time rehe}, with respect to the addressed potentials we approximately set $\tau_f$ to a magnitude between $\sim 5\cdot10^{17}$ {GeV}$^{-1}$ and $\sim 3\cdot10^{18}$ {GeV}$^{-1}$. For this work, we fix $\tau_f=5\cdot10^{17}$ {GeV}$^{-1}$.

Within this framework, the \emph{out}-mode expansion can be then written in terms of $\tilde f_k(\tau)=f_k(\tau)/a(\tau)$
\begin{equation}\label{fourierquant out}
    \hat{\delta\phi}_{out}=\frac{1}{(2\pi)^{3/2}}\int{d^3k\left(\hat{b}_k\tilde f_k(\tau)e^{i{\bf k\cdot x}}+\hat{b}_k^\dagger\tilde f_k^*(\tau)e^{-i{\bf k\cdot x}}\right)},
\end{equation}
with $\hat{b}_k^\dagger$ and $\hat{b}_k$ the creation and annihilation operators.
These operators refer to the final state in the \emph{out} region, namely where the adiabatic condition of Eq. \eqref{adiab} is recovered.
They are defined through the Bogoljubov transformation, which connects the initial state ladder operators, $\hat{a}_k,\,\hat{a}_k^\dagger$, with the aforementioned final state ones via the Bogoljubov coefficients $\alpha_k,\,\beta_k$ \cite{Parker}
\begin{equation}\label{out ladder}
\hat{b}_k=\alpha_k\hat{a}_k+\beta_k^*\hat{a}_k^\dagger.
\end{equation}
Equivalently, the Bogoljubov coefficients transform the \emph{in-}state to the \emph{out}-state of the quantum field.

It can be shown that the Bogoljubov coefficients describing the transition of inflaton fluctuations from slow-roll to the reheating stage can be expressed as
\begin{subequations}
    \begin{align}\label{alpha}
    \alpha_k&=i\left[\delta\phi_k^\prime(\tau_f)\tilde f_k^*(\tau_f)-\delta\phi_k(\tau_f)\tilde f^{\prime*}_k(\tau_f)\right],\\
    \beta_k&=-i\left[\delta\phi_k^\prime(\tau_f)\tilde f_k(\tau_f)-\delta\phi_k(\tau_f)\tilde f^{\prime}_k(\tau_f)\right].\label{beta}
    \end{align}
\end{subequations}


\section{Particle production during and after inflation}\label{Sect 4}

In the literature, geometric particle production, or more broadly, particle production \emph{during inflation} are generally ignored as due to the above-discussed issue related to properly identify particle states during the non-adiabatic de Sitter-like expansion. Furthermore, a pure de Sitter spacetime would result in vanishing Bogoljubov coefficients \cite{Birrell:1982ix} and, in general, geometric production is seen as a second order expansion with respect to the gravitational particle production.

However, when considering the transition to reheating, gravitational production from the inflaton becomes significant and particle densities can be computed unambiguously.
Furthermore, the presence of inhomogenities associated with inflationary fluctuations typically gives rise to additional particles via perturbative processes. In so doing, we discuss below the gravitational and geometric particle productions and distinguish their main features, showing that a possible comparison is instead possible even during inflationary stage.

\subsection{Gravitational production}\label{sect 4}

We have defined the cosmological framework describing a single-field inflationary epoch and established coherent boundary conditions.
Now, we aim to investigate primordial particle creation processes within a perturbative approach that involves spacetime inhomogeneities \cite{cespedes,friemanpart}.

Specifically, we encode the inhomogeneities in the perturbed term of the metric $g_{\mu\nu}$, as defined in Eq. \eqref{perturbed metric}. 
We first proceed in the Jordan frame, while in Sect. \ref{sec 4.3} we properly address the frame issue through the mechanisms of particle production.

According to this picture, particle production is investigated through the $S$ matrix formalism\footnote{ We clarify that by the $S$ matrix, we refer to the formalism introduced for handling scattering processes and to the matrix elements derived from the $\hat{S}$ operator. Therefore, in our notation, we omit the $\hat{}$ symbol when discussing about the $S$ matrix.}. In particular, we can write the final quantum state up to fourth order as
\begin{align}
    \label{finstat}
&\lvert \Psi \rangle = \hat{S} \lvert 0 \rangle = \mathcal{N} \bigg( \lvert 0 \rangle + \frac{1}{2} \int d^3{p_1}d^3{p_2}\mathcal{C}_{{\bf p}_1,{\bf p}_2} \lvert {{\bf p}_1},{{\bf p}_2} \rangle  \notag \\
&+\frac{1}{24}\int d^3{q_1}d^3{q_2}d^3{q_3}d^3{q_4} \mathcal{C}_{{\bf q}_1,{\bf q}_2,{\bf q}_3,{\bf q}_4} \lvert {{\bf q}_1},{{\bf q}_2},{{\bf q}_3},{{\bf q}_4} \rangle \bigg),
\end{align}
where $\lvert 0 \rangle$ is the Bunch-Davies vacuum, $\mathcal{N}$ a normalization constant given by the condition $\langle\Psi\vert\Psi\rangle$=1, and $\mathcal{C}_{{\bf p}_1,{\bf p}_2},\,\mathcal{C}_{{\bf q}_1,{\bf q}_2,{\bf q}_3,{\bf q}_4}$ are the probability amplitudes for the production of pairs and quartets of particles, respectively.
Hence, we can compute the total number of particles as
\begin{equation}\label{totnum}
    N=\langle\Psi\vert\hat{N}\vert\Psi\rangle,
\end{equation}
where we introduce the number operator 
\begin{equation}
   \hat N =\frac{1}{(2\pi a)^3}\int d^3k\, \hat{b}_k^\dagger \hat{b}_k ,
\end{equation}
with $\hat{b}_k^\dagger$ and $\hat{b}_k$ the creation and annihilation operators in the \emph{out} region, as shown by Eq. \eqref{out ladder}.
Using these definitions, we can explicitly calculate Eq. \eqref{totnum} by exploiting the perturbative expansion of Eq. \eqref{finstat}.
Accordingly, perturbative particle production from inflationary inhomogeneities inevitably depends on the choice of the potential driving the inflationary epoch. We will show that, within the above-introduced potentials of Eqs. \eqref{larpot} -- \eqref{nat pot}, we can focus on the direct products between the particles pairs, quartets and vacuum state.
We then notice that:
\begin{widetext}
    \begin{align}
    N^{(2)}=&\ {}^{(2)}\langle\Psi|\hat N|\Psi\rangle^{(2)}\\
    =&\ \frac{\mathcal{N}^2}{4\left(2\pi a(\tau)  \right)^{3}} \int d^3p_1\  d^3p_2\  d^3k\ \lvert \mathcal{C}_{{\bf p}_1,{\bf p}_2}\rvert^2 \langle {\bf p_1p_2} \lvert \hat b^\dagger_k\hat b_k \rvert {\bf p_1p_2}\rangle\\
    =&\ \frac{\mathcal{N}^2}{4\left(2\pi a(\tau)  \right)^{3}} \int d^3p_1\  d^3p_2\  d^3k\ \lvert \mathcal{C}_{{\bf p}_1,{\bf p}_2}\rvert^2 \langle {\bf p_1p_2} \lvert \ |\alpha_k|^2\hat a^\dagger_k\hat a_k+|\beta_k|^2\hat a_k\hat a_k^\dagger\rvert {\bf p_1p_2}\rangle\\
    =&\ \frac{\mathcal{N}^2}{4\left(2\pi a(\tau)  \right)^{3}} \int d^3p_1\  d^3p_2\  d^3k\ \lvert \mathcal{C}_{{\bf p}_1,{\bf p}_2}\rvert^2 \left( \left( |\alpha_k|^2+|\beta_k|^2\right))\left(\delta^{(3)}({\bf k-p_1})+\delta^{(3)}({\bf k-p_2})\right)\right)\\
    =&\ \frac{\mathcal{N}^2}{2\left(2\pi a(\tau)  \right)^{3}} \int d^3p_1\  d^3p_2\   \lvert \mathcal{C}_{{\bf p}_1,{\bf p}_2}  \rvert^2 \left( 1+ \lvert \beta_{p_1} \rvert^2+ \lvert \beta_{p_2} \rvert^2 \right).\label{num of couples}
\end{align}
\end{widetext}
Similarly, it can be shown that:
\begin{align}
N^{(0)}(\tau)=& \frac{\mathcal{N}^2}{\left(2\pi a(\tau)  \right)^{3}} \int d^3k\, \lvert \beta_{k} \rvert^2, \label{grav part prod} \\
    N^{(4)}(\tau)=&\frac{\left(\mathcal{N}/6\right)^2}{\left(4\pi a(\tau)  \right)^{3}} \int d^3q_1\  d^3q_2\ d^3q_3\  d^3q_4\ \notag \\ 
    &\times \lvert \mathcal{C}_{{\bf q}_1,{\bf q}_2,{\bf q}_3,{\bf q}_4}  \rvert^2  \left( 2+ \sum_{i=1}^{4}\lvert \beta_{q_i} \rvert^2 \right)\label{num of quartets}.
\end{align}

The aforementioned particles production is classified into two different regimes:
\begin{itemize}
    \item[-] the particle number in Eq. \eqref{grav part prod} is referred to as \emph{gravitational particle production}, see \emph{e.g.} Refs. \cite{fullingparker,Ford:2021syk,ford}; this mechanism is well-known and widely studied in the literature, since it is typically considered the dominant particle production process arising from purely gravitational effects. It depends on Bogoljubov transformations, and involves particle-antiparticle pairs, which are produced from vacuum as consequence of the non-adiabatic universe expansion;
    \item[-] the particle number in Eqs. \eqref{num of couples}, \eqref{num of quartets} represent the contributions associated with \emph{geometric particle production} \cite{friemanpart,cespedes}, which we highlighted at second and fourth order, respectively. The presence of spacetime inhomogeneities in an expanding universe leads to this geometric mechanism of particle production that is not subdominant \emph{a priori} \cite{Belfiglio:2022cnd,Belfiglio:2022yvs,Belfiglio:2023eqi,Belfiglio:2023rxb,belfiglio2023superhorizon,Belfiglio:2024xqt}. Its origin is again of gravitational form, despite its nature is significantly different from the non-perturbative processes described by Eq. \eqref{grav part prod}. In particular, the presence of a Bogoljubov-independent term provides an additional creation of particles, and in general, the resulting particles do not need to assume opposite momenta, thus introducing mode-mixing in particle production processes.

\end{itemize}


\subsection{Geometric  production}\label{sect 4.1}

We now delve into the geometric mechanism of particle production, to highlight its main features within the inflationary scenario.

To do so, we start from the zeroth-order energy-momentum tensor related to the inflation fluctuactions, 
\begin{align}
    T_{\mu \nu}^{(0)}=& \partial_{\mu} \delta \phi\  \partial_{\nu}\delta \phi- g_{\mu \nu}^{(0)} \left[ \frac{1}{2}g^{\rho \sigma}_{(0)}\  \partial_{\rho} \delta \phi\  \partial_{\sigma} \delta \phi - V(\delta \phi)  \right] \notag \\ 
&+ \xi \left[ g_{\mu \nu}^{(0)} \nabla^\rho \nabla_\rho- \nabla_{\mu} \partial_{\nu}+R_{\mu \nu}^{(0)}-\frac{1}{2} R^{(0)} g_{\mu \nu}^{(0)}   \right] (\delta \phi)^2,
\end{align}
where $R^{(0)},\,R^{(0)}_{\mu\nu}$ represent, respectively, the zeroth-order Ricci scalar and tensor.
Consequently, we can write the first-order interaction Lagrangian density between the perturbations and the inflaton field fluctuations as
\begin{equation} \label{intlagg}
    \mathcal{L}_{I}=-\frac{1}{2}\sqrt{-g^{(0)}}\delta g^{\mu\nu}T^{\left(0\right)}_{\mu\nu}.
\end{equation}
In particular, if such Lagrangian satisfies the property $\mathcal{L}_I=-\mathcal{H}_I$, which holds for the above-presented inflationary potentials, we can exploit Dyson's expansion \cite{dyson} at first order to write
\begin{equation}
    \hat{S} \simeq 1 + i {T} \int d^4x \mathcal{L}_I,
\end{equation}
where $T$ is the time ordering symbol.

We emphasize that defining the $S$ matrix in a context of curved spacetime is typically not trivial \cite{Wald:1979kp,Audretsch:1985vy}.
Specifically, ensuring the adiabaticity condition, as stated in Eq. \eqref{adiab}, in the initial and final states, is crucial in order to properly deal with particle states when computing probability amplitudes.

Particle creation through the $S$ matrix is, by definition, provided by second quantization of the fields. 
The fluctuations modes can be quantized as in Eq. \eqref{fourierquant}, but, in order to explicitly write the mode functions, we need to specify the inflationary potential first.

Further, the computation of probability amplitudes is usually troublesome for potentials that are not polynomial in form.

To circumvent this, a possible solution is to approximate our potentials, when possible, with an even power series up to the fourth order, under the form 
\begin{equation}\label{pot approx}
    V(\phi)\simeq v_0+v_2(c+\phi)^2+v_4(c+\phi)^4,
\end{equation}
with free numerical parameters $v_0,\,v_2,\,v_4,\,c$ to determine. 
In this respect, during inflation, we evaluate particle production within a limited interval, from the horizon exit corresponding to the pivot scale up to the end of inflation. 
Hence, the aforementioned parameters are determined by fitting the power series within this interval.
In particular, we focus on fluctuations, so by shifting the field by a constant value $c$, the inflaton fluctuations remain unaffected, resulting in
\begin{equation}
    V(\delta\phi)\simeq v_0+v_2\cdot(\delta\phi)^2+v_4\cdot(\delta\phi)^4.
\end{equation}

Within this approximation, we can compute the probability amplitude to obtain pairs and quartets of particles by considering the $S$ matrix up to first order in Dyson's expansion. 
Specifically, the pair production probability is recovered by
\begin{align} \label{compact}
\mathcal{C}_{{\bf p}_1,{\bf p}_2} & \equiv \langle {\bf p}_1,{\bf p}_2 \lvert \hat S \rvert 0 \rangle \notag \\ 
& = -\frac{i}{2} \int d^4x\  2a^2\big(A_0({\bf x}, \tau)
+A_1({\bf x},\tau) \notag \\
&\ \ \ \ \ \ \ \ \ \ \ \ \ \ \ \ \ \ \ \ \ \ \ \ \ \ +A_2({\bf x},\tau)+A_3({\bf x},\tau) \big),
\end{align}
where we define $A_0({\bf x},\tau)$ and $A_i({\bf x},\tau)$ for the time and spatial components, respectively, as
\begin{align} \label{Atime}
A_0({\bf x},\tau)=  \Psi \bigg[&\partial_0 \delta \phi_{p_1}^*\  \partial_0 \delta \phi_{p_2}^* \notag \\
&- \big( \frac{1}{2}\eta^{\rho \sigma} \partial_\rho \delta \phi_{p_1}^* \ \partial_\sigma \delta \phi_{p_2}^*-a^2 \tilde V_2(\delta \phi) \big) \notag \\
    & \xi \bigg(\partial_0 \partial_0-3\frac{a^\prime}{a}\partial_0 -\eta^{\rho \sigma}\partial_\rho \partial_\sigma \notag \\
    &\ \ \ \ \ \ \ +3\left( \frac{a^\prime}{a} \bigg)^2   \right) \delta \phi_{p_1}^* \delta\phi_{p_2}^* \bigg] e^{-i({\bf p_1}+{\bf p_2})\cdot {\bf x}}\notag\\
\end{align}
and 
\begin{align} \label{Aspace}
    A_i({\bf x},\tau)= \Psi \bigg[&\partial_i \delta \phi_{p_1}^*\  \partial_i \delta \phi_{p_2}^*\notag \\
    & + \big( \frac{1}{2}\eta^{\rho \sigma} \partial_\rho \delta \phi_{p_1}^* \ \partial_\sigma \delta \phi_{p_2}^*-a^2 \tilde V_2(\delta \phi) \big)\notag \\
    &+ \xi \bigg(\partial_i \partial_i+\frac{a^\prime}{a}\partial_0+\eta^{\rho \sigma}\partial_\rho \partial_\sigma - \frac{2a^{\prime \prime}}{a}\notag \\
    &\ \ \ \ \ \ \ \ -\left( \frac{a^\prime}{a} \right)^2  \bigg) \delta \phi_{p1}^* \delta\phi_{p2}^* \bigg] e^{-i({\bf p_1}+{\bf p_2})\cdot {\bf x}},\notag\\
\end{align}
with $\tilde V_2(\delta\phi)$ representing the potential contribution.
Further, to explicitly write $\tilde V_2(\delta\phi)$, we recall the Wick's theorem that yields \cite{wick}
\begin{equation}
    \tilde V_2(\delta\phi)=\left(v_2+12v_4\Delta\right)\delta\phi_{p1}\delta\phi_{p2},
\end{equation}
with $\Delta$ the free propagator\footnote{For a more rigorous and specific analyses of the Feynmann propagator in curved spacetime, see \emph{e.g.} Ref. \cite{Feynprop}.}.

The free propagator is, by definition, a divergent quantity, requiring regularization procedures. 

More generally, to properly tame the Feynman propagator and to naturally remove the divergent quantities, we might consider the renormalization group\footnote{Such renormalization procedure is not straightforward in curved spacetime, thus we leave these considerations for future works.} \cite{Bunch:1978yq}.
However, since regularization consists in an approximation to face these infinities trough cutoffs, we impose an UV cutoff on the momenta, setting the maximum momentum, $k_{max}$, namely
\begin{equation}
    k_{max}=k_f=a(\tau_f) H_I. 
\end{equation}
In other words, we focus on modes that spill out the horizon by the end of inflation, discarding those that remain inside the horizon during reheating. The latter  are then sensitive to the reheating dynamics and thermalization processes on shorter timescales \cite{Boya}. 
Consequently, the choice over momenta, despite physically reasonable, clearly represents a first approximation, that ignores the coupling of the inflaton to other quantum fields during reheating.

Moving to particle quartets, we have a similar expression for the probability amplitude, which now reads
\begin{align} \label{compactB}
\mathcal{C}_{{\bf q}_1,{\bf q}_2,{\bf q}_3,{\bf q}_4} & \equiv \langle {\bf q}_1,{\bf q}_2,{\bf q}_3,{\bf q}_4 \lvert \hat S \rvert 0 \rangle \notag \\ 
& = -\frac{i}{2} \int d^4x\  2a^2\big(B_0({\bf x}, \tau)
+B_1({\bf x},\tau) \notag \\
&\ \ \ \ \ \ \ \ \ \ \ \ \ \ \ \ \ \ \ \ \ \ \ \ \ \ +B_2({\bf x},\tau)+B_3({\bf x},\tau) \big),
\end{align}
where we introduce
\begin{align} \label{B}
B_0({\bf x},\tau)=-B_i({\bf x},\tau)
=\Psi a^2 \tilde V_4(\delta \phi)e^{-i({\bf q_1}+{\bf q_2}+{\bf q_3}+{\bf q_4})\cdot {\bf x}},
\end{align}
and
\begin{equation}
    \tilde V_4(\delta\phi)=12v_4\delta\phi_{q1}\delta\phi_{q2}\delta\phi_{q3}\delta\phi_{q4}.
\end{equation}

Bearing this general framework in mind, in the next section we will explore particle production for the specific models,  previously introduced, examining both frames and regimes.


\section{Comparing geometric and gravitational particle production}\label{Sect 5}

Our aim is to compare geometric and gravitational particle production and to establish which mechanism is dominant, and under which assumptions. Extending this analysis to both frames, we can also provide a further test for frame equivalence.

In this respect, we first deal with the Jordan frame and, motivated by the positive outcomes of the inflationary paradigm, we study in detail the corresponding potentials introduced in  Eqs. \eqref{larpot} -- \eqref{nat pot}.

To analyze these models, we begin by expressing all dynamical quantities in terms of conformal time, denoted by $\tau$. Assuming a quasi-de Sitter evolution, as described by Eq. \eqref{desitter}, we evaluate the Friedmann equation, Eq. \eqref{desitter}, and the continuity equation, Eq. \eqref{Jordan dyn}, at the onset of inflation, which allows us to determine the value of $H_I$.

Further, as previously mentioned, we fix the end time of inflation at  $\tau_f=5\cdot10^{17}$ GeV$^{-1}$, and approximate the slow-roll parameter as $\epsilon=0.003$. With these values, we can compute the initial time $\tau_0$ by imposing the e-folding number as $N=70$, thus finding
\begin{equation}
    \tau_0=-e^{70/(1+\epsilon)}\tau_f+2\tau_f.
\end{equation}

Since the initial and final times are determined \emph{a priori} for all potentials, we observe slight modifications in the other dynamical parameters.  
In particular, from Eq. \eqref{time rehe}, we compute the reheating e-folding number as
\begin{equation}
    N_{re}=\ln{(H_IH_{rad})\tau_f^2}.
\end{equation}
Consequently, by imposing continuity between epochs, we obtain
\begin{equation}
    H_{re}=\sqrt{\frac{4H_{rad}}{\tau_fe^{N_{re}}}}.
\end{equation}
We also introduce the final time for the reheating phase, corresponding to the onset of the radiation-dominated era and denoted by $\tau_r=\tau_fe^{N_{re}}$, the time when the adiabaticity condition is recovered, Eq. \eqref{adiab}, denoted as $\tau_{ad}$, and finally, the time when the modes exit the horizon at the pivot scale, namely $\tau_*\simeq-1/k_*=-7.8\cdot10^{40}$ GeV$^{-1}$.
Clearly, the following conditions are verified:
\begin{equation}
    \tau_0<\tau_*<\tau_f<\tau_{ad}<\tau_r.
\end{equation}

Regarding the background evolution of the inflaton field, it is typically not possible to find an analytical expression $\phi(\tau)$ for the inflationary potentials under consideration. Thus, we adopt the following ansatz
\begin{equation}
    \phi(\tau^\prime)=c_1\vert\tau^\prime\vert^{c_2}+c_3,
\end{equation}
where $\tau^\prime$ is the conformal time rescaled as in the scale factor, Eq. \eqref{desitter}, and $c_1,\,c_2$ and $c_3$\footnote{In Tab. \ref{tab part} we report all the numerical values for the free parameters of the ansatz. Specifically, the constant $c_1$ dimensionally reads GeV$^{(c_2+1)}$, while $c_2$ is a pure number and $c_3$ scales as GeV.} are constants to be determined.
By imposing the initial and final conditions, as outlined in Tab. \ref{tab nonmin}, it is straightforward to compute the values of the three parameters.

\begin{figure*}[ht]
  \centering
  \textbf{a)}\hspace{0.mm}
    \includegraphics[width=.465\linewidth]{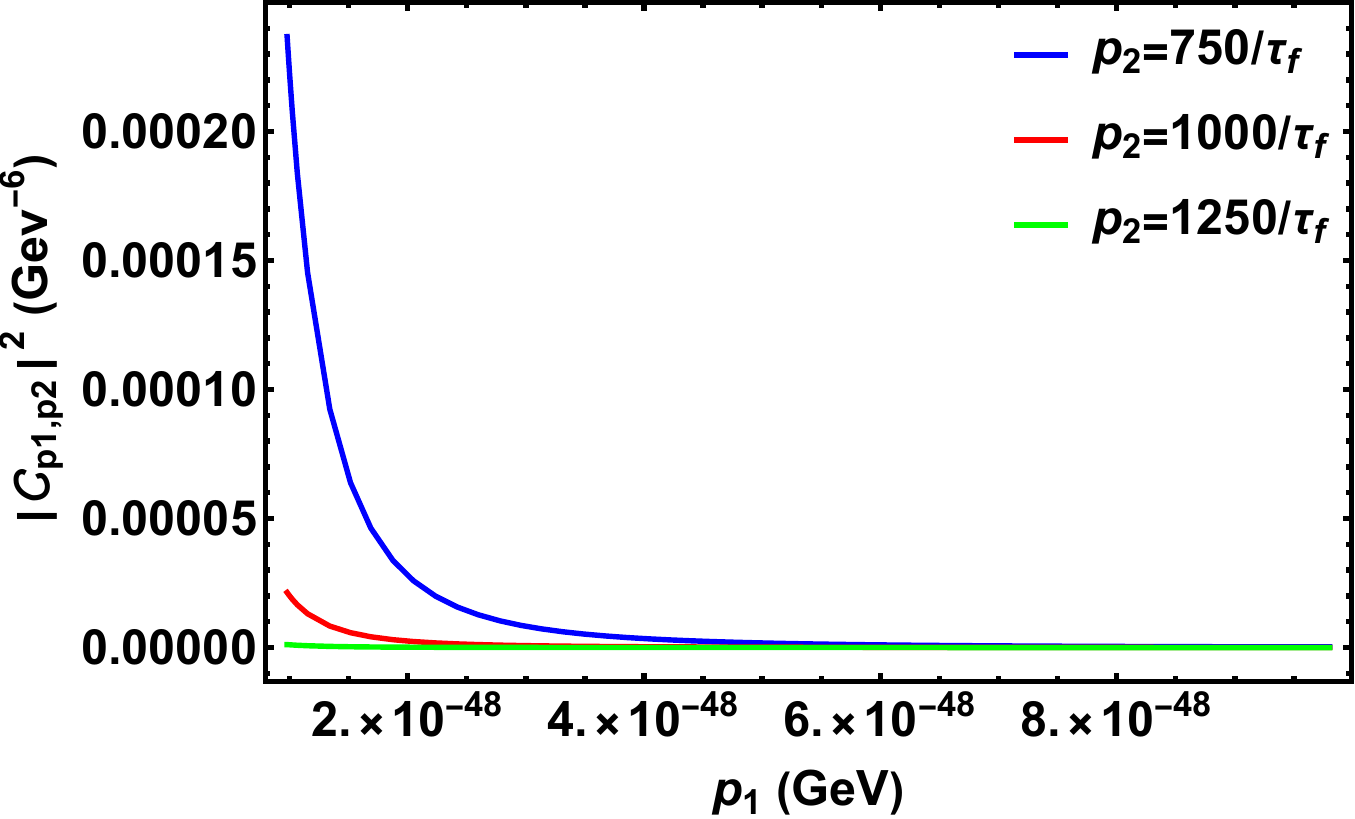}
  \hspace{3mm}
  \textbf{b)}\hspace{1.5mm}
\includegraphics[width=.435\linewidth]{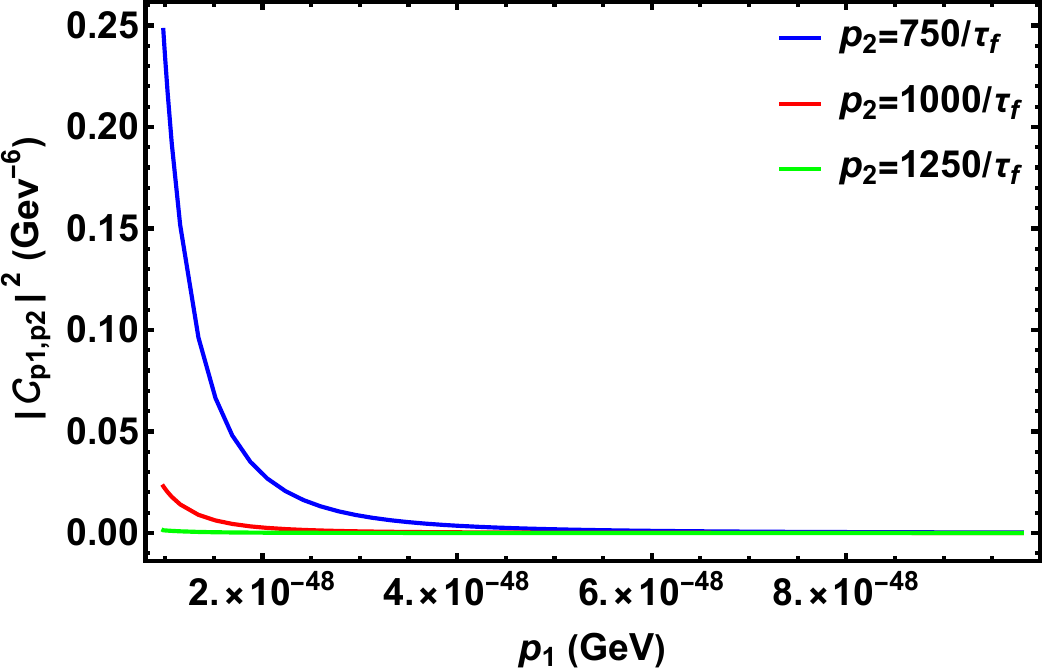}
  \hspace{0mm}
  \textbf{c)}\hspace{5mm}
    \includegraphics[width=.45\linewidth]{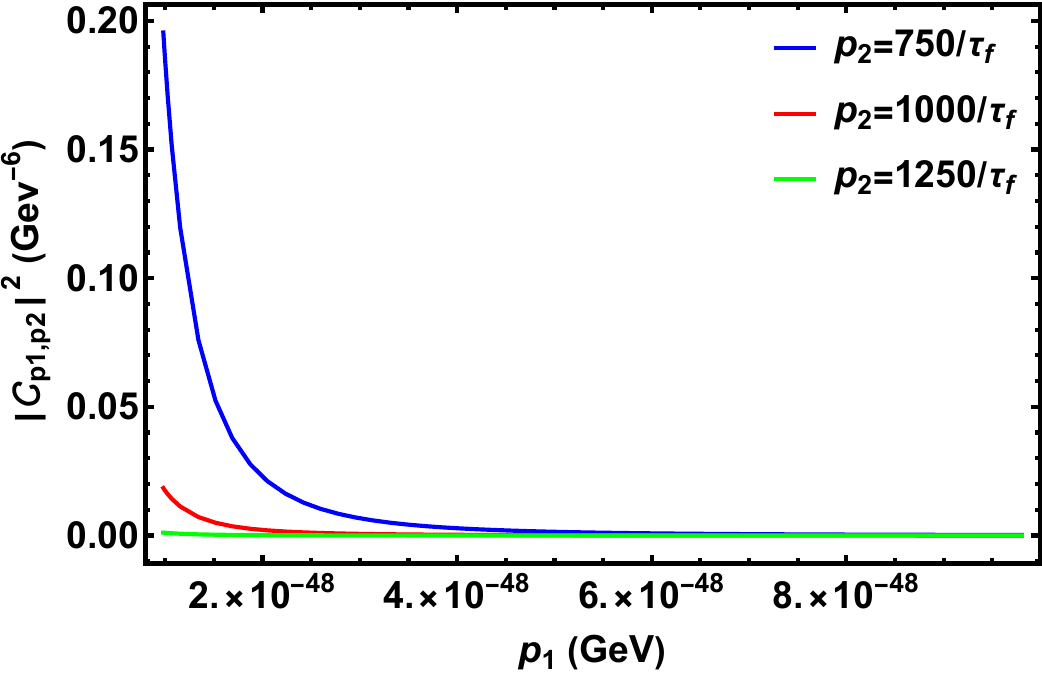}
  \hspace{0mm}
    \textbf{d)}\hspace{.mm}
    \includegraphics[width=.46\linewidth]{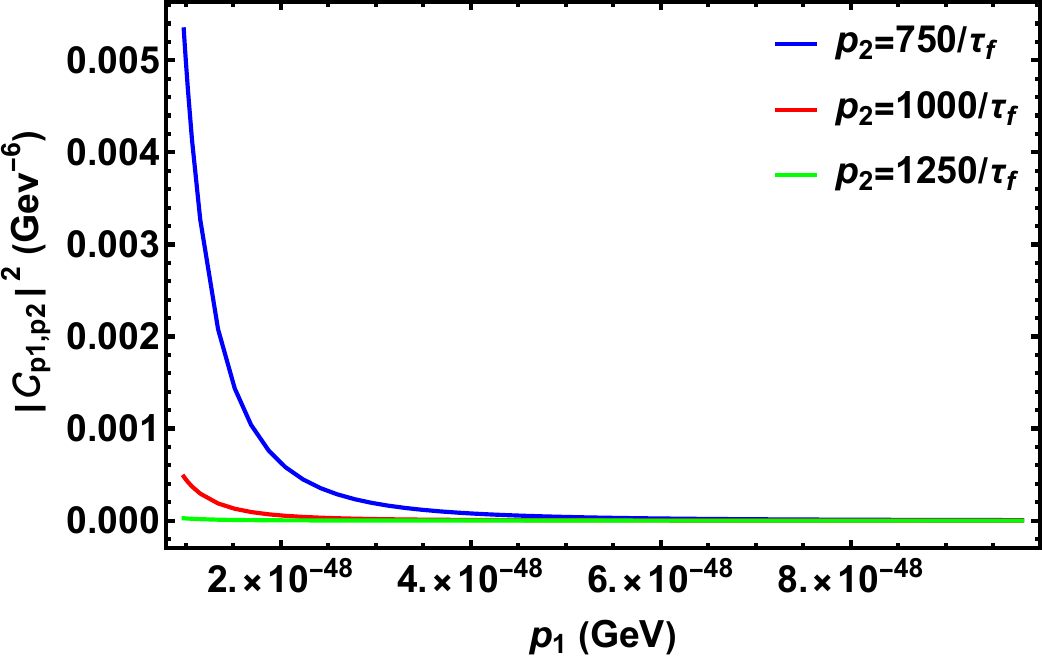}  \hspace{0mm}
  \caption{Ratio between geometric and gravitational pairs production densities, assuming respectively: \textbf{a)} the Starobinsky potential $V_S(h)$, from Eq. \eqref{star}; \textbf{b)} the $E-$model $V_E(h)$, from Eq. \eqref{E pot}, with $\alpha^E=10$; \textbf{c)} the $T-$model $V^E_T(h)$, from Eq. \eqref{T pot}, with $\alpha^T=0.1$; \textbf{d)} natural inflation potential $V_N(h)$, from Eq. \eqref{nat pot}, with $f=1.5M_{Pl}$.}
  \label{fig part pairs}
\end{figure*}

\begin{figure*}[ht]
  \centering
  \textbf{a)}\hspace{0.mm}
    \includegraphics[width=.45\linewidth]{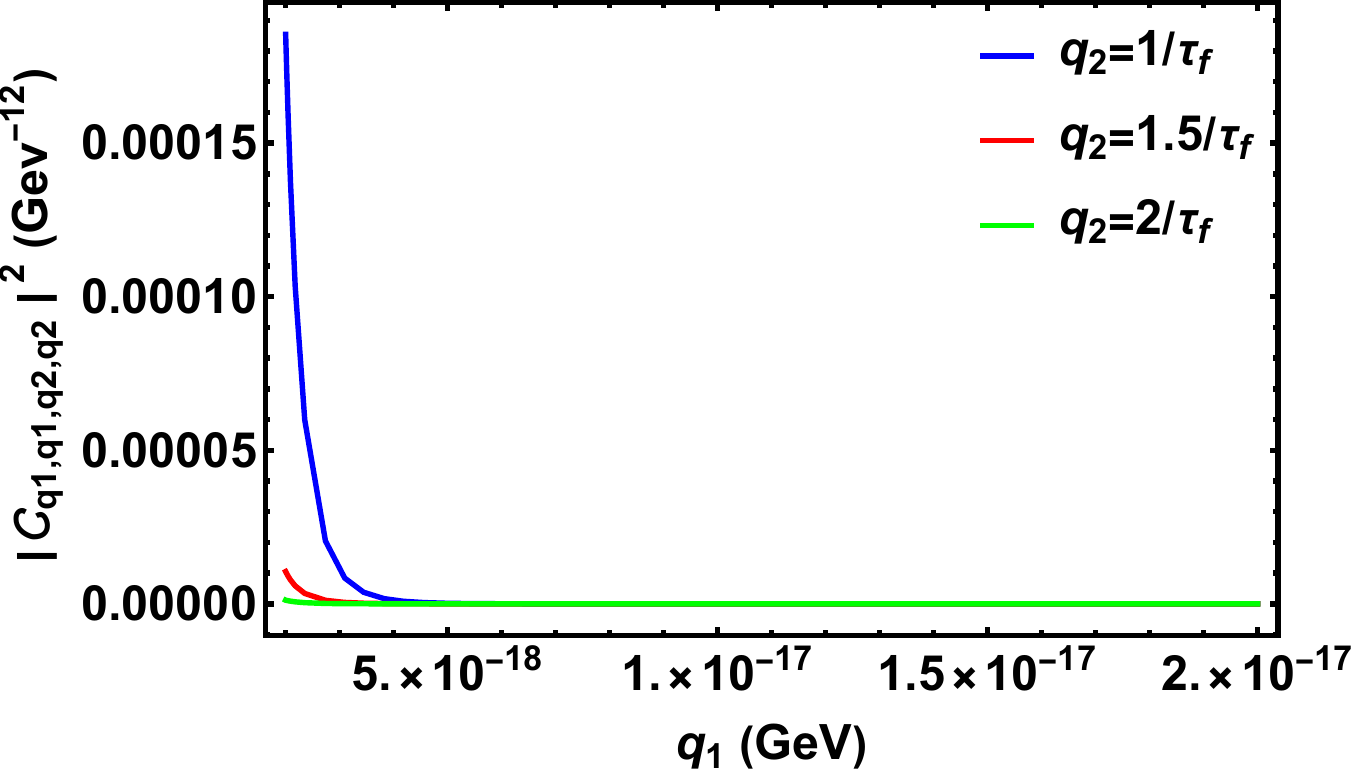}
  \hspace{1.5mm}
  \textbf{b)}\hspace{0mm}
\includegraphics[width=.46\linewidth]{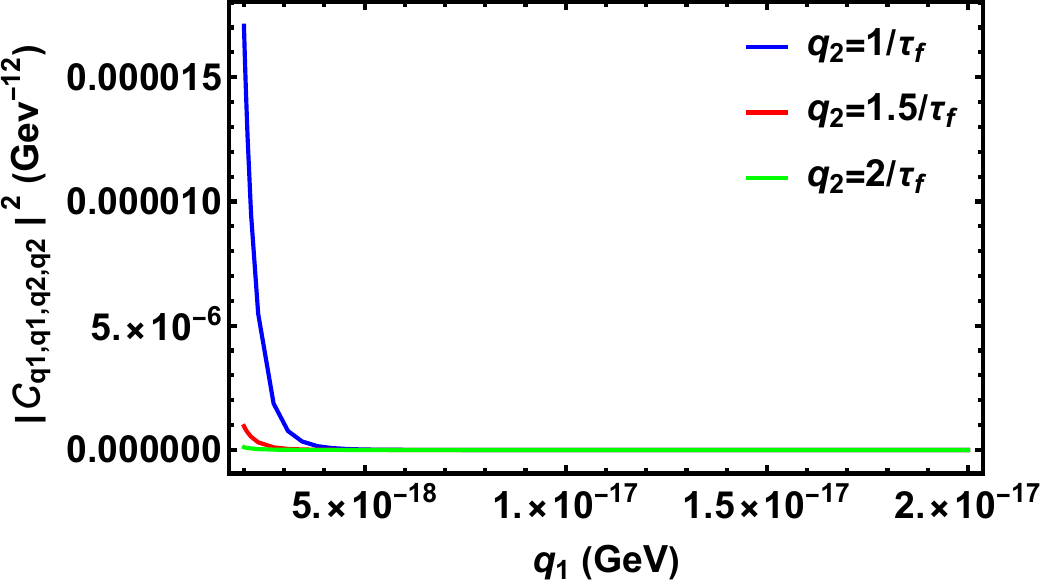}
  \hspace{0.mm}
  \textbf{c)}\hspace{3.5mm}
    \includegraphics[width=.44\linewidth]{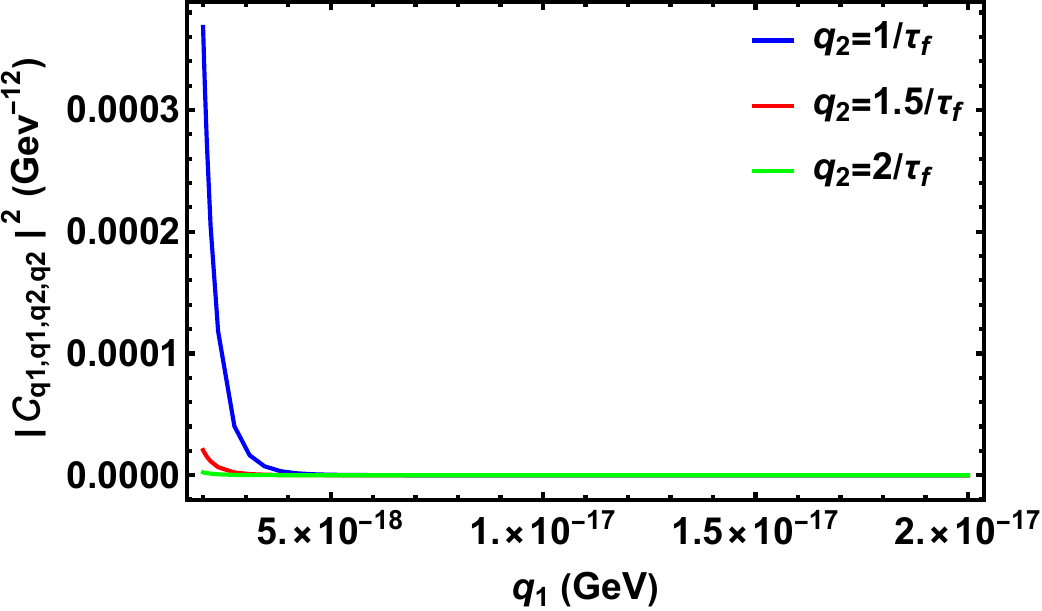}
  \hspace{1.5mm}
    \textbf{d)}\hspace{1.5mm}
    \includegraphics[width=.455\linewidth]{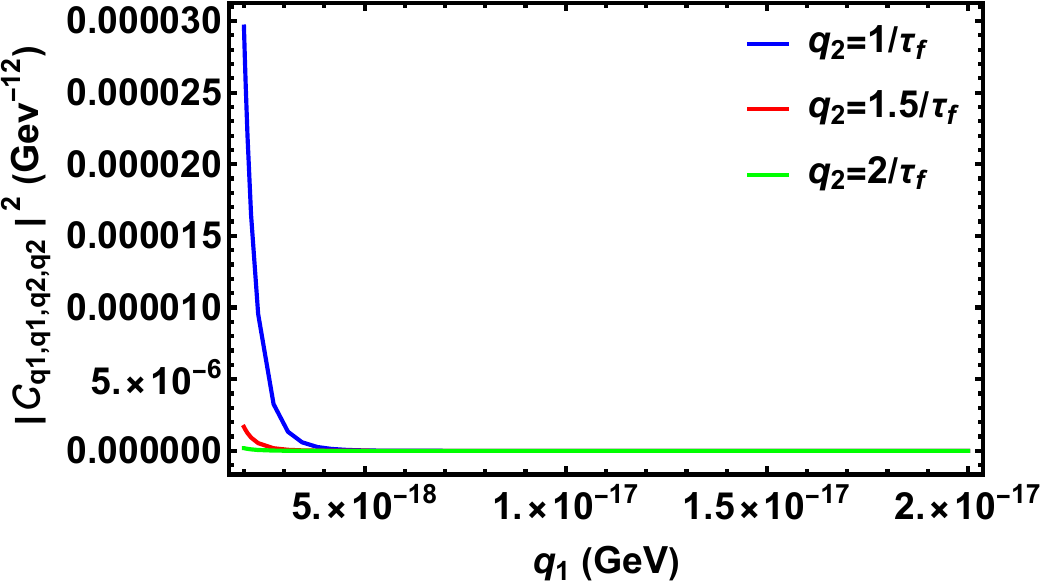}  \hspace{0mm}
  \caption{Ratio between geometric and gravitational quartets production densities, respectively: \textbf{a)}the Starobinsky potential $V_S(h)$, from Eq. \eqref{star}; \textbf{b)} the $E-$model $V_E(h)$, from Eq. \eqref{E pot}, with $\alpha^E=10$; \textbf{c)} the $T-$model $V^E_T(h)$, from Eq. \eqref{T pot}, with $\alpha^T=0.1$; \textbf{d)} natural inflation potential $V_N(h)$, from Eq. \eqref{nat pot}, with $f=1.5M_{Pl}$.}
  \label{fig part quartets}
\end{figure*}

This expression for the background evolution is needed to compute the scalar perturbation potential $\Psi$ and the inflaton fluctuation $\delta\phi$. However, because of the numerical complexity of the involved equation, we simplify by reducing the number of functions. Specifically, we substitute each functional of the background scalar field with the corresponding average
\begin{equation}\label{average}
    F[\phi(\tau)]\rightarrow\frac{1}{\tau_f-\tau_0}\int_{\tau_0}^{\tau_f}d\tau F[\phi(\tau)].
\end{equation}

Finally, we can explicitly write $\nu$, Eq. \eqref{nu}, and so derive the fluctuations $\delta\phi$, Eqs. \eqref{subfluc} and \eqref{supfluc}. 
By solving Eq. \eqref{psipsi} and imposing the super-Hubble limit of Eq. \eqref{psipsi0}, namely $\Psi_k\propto\delta\phi_k$, we obtain the perturbation potential
\begin{equation}
    \Psi_k(\tau)= C(k)\tau^\zeta H_{\nu_\Psi}^{(1)}(-k\tau),
\end{equation}
where $C(k)$ is an integration constant that could, in principle, depend on the modes, and $\zeta$ is an adimensional constant. 
Observational constraints suggest that the perturbation at the pivot scale satisfies $|\Psi_{k_*}(\tau_*)|\simeq10^{-5}$, providing a bound for the value of $C(k)$ \cite{nohair}.
We emphasize that, once again, to compute the amplitude of probability, we consider the average of the perturbation potential in analogy to Eq. \eqref{average}.

The potential, as already discussed, is approximated by modeling it as an even power series up to fourth order, Eq. \eqref{pot approx}. We remark that, despite the fitted potentials present a sign difference between the second and fourth order, we are not dealing with symmetry breaking potentials because they just represent numerical approximations of the starting models, and so they do not assume any further physical meaning here.

\begin{table}[h]
  \centering
  \begin{tabular}{c|c|c|c|c}
    \hline\hline
     & $c_1\ [GeV^{\alpha}]\cdot 10^{19}$& $v_0\,[GeV^4]\cdot10^{63}$ & $\nu$ &$C\ [GeV^{\beta}]\cdot 10^{-26}$ \\
    \hline
    \hline
    $V_S$ & $\,-8.32$& $-0.76$ & $1.501$  &  $27.0 $ \\
    
    $V_E$ & $\,-12.6$& $-0.04$ & $1.508$ &  $49.1$  \\
    
    $V_T$ & $\,-7.45$& $-17.4$ & $1.509$ & $64.9$ \\
    
    $V_N$ & $\quad10.5$& $\quad24.1$  & $1.504$  & $23.9$\\  
     \hline\hline
     & $c_2 \cdot 10^{-1}$ & $v_2\,[GeV^2]\cdot10^{26}$& $\nu_\Psi$& $N_{re}$  \\
    \hline
    \hline
    $V_S$ & $-0.22$  & $\quad0.14$ & $0.525$ & $10.8$  \\
    
    $V_E$ & $-0.28$  & $\quad0.03$ & $0.531$ & $10.4$  \\
    
    $V_T$ & $-0.31$  & $\quad0.22$ & $0.534$ & $11.0$  \\
    
    $V_N$ & $-0.21$  & $\,-0.10$ & $0.524$ & $11.2$ \\ 
    \hline
    \hline
    &$c_3\ [GeV]\cdot 10^{19}$ & $v_4\cdot 10^{-14}$  & $\zeta$& $H_{re}\,[GeV]\cdot10^{-33}$ \\
    \hline
    \hline
    $V_S$ & $3.60$  & $\,-0.73$ & $0.481$& $1.25$  \\
    
    $V_E$ & $4.18$ & $\,-0.06$ & $0.475$& $1.58$   \\
    
    $V_T$ & $2.30$  & $\,-0.05$ & $0.472$& $1.16$  \\
    
    $V_N$ & $1.08$ & $\quad0.11$ & $0.482$& $1.06$ \\  
    \hline
    \hline
  \end{tabular}
  \caption{Resuming table of all the parameters computed in the particle production framework. Specifically, $\alpha=c_2+1$ and $\beta=\frac{3}{2}+\zeta$.}
  \label{tab part}
\end{table}

All the parameters and quantities introduced above are listed in Tab. \ref{tab part}.

Accordingly, we can properly compute the particle number densities arising from gravitational and geometric effects.
We notice that the total number densities, Eqs. \eqref{grav part prod}, \eqref{num of couples}, \eqref{num of quartets}, strongly depend on the Bogoljubov coefficients. 
As we already mentioned, these parameters vanish in the limit of a pure de Sitter spacetime. Thus, to qualitatively understand particle production during inflation, we may compute the amplitudes $\lvert \mathcal{C}_{{\bf p}_1,{\bf p}_2}  \rvert^2$ and $\lvert \mathcal{C}_{{\bf q}_1,{\bf q}_2,{\bf q}_3,{\bf q}_4}  \rvert^2$, as only the pure geometric contribute survives.

However, in a realistic scenario which includes transition to reheating, where adiabaticity is recovered, it can be shown from Eq. \eqref{beta} that the Bogoljubov coefficient $\beta$ is much larger than 1 and, specifically, it is much larger for super-Hubble modes with respect to sub-Hubble ones. 
Thus, referring to Eqs. \eqref{num of couples}, \eqref{num of quartets}, we see that the ratio between gravitational and geometric contributions simply behaves as $\lvert \mathcal{C}_{{\bf p}_1,{\bf p}_2}  \rvert^2$ and $\lvert \mathcal{C}_{{\bf q}_1,{\bf q}_2,{\bf q}_3,{\bf q}_4}  \rvert^2$. 
This implies that the geometric contribution to the total particle density can be inferred from the above-mentioned geometric probability amplitudes, as depicted in Figs. \ref{fig part pairs}, \ref{fig part quartets}.

According to Eqs. \eqref{compact} and \eqref{compactB}, we expect the particle rate to be higher for positive couplings and lower for negative couplings. 
In the weakly interacting limit, corrections due to curvature effects are typically subdominant within the potential contribution, so, for simplicity, we focus only on positive coupling constants $\xi$ in our analysis.

When computing the probability amplitudes, we need to focus on modes that result in sufficiently small perturbations, namely $\lvert \Psi_k \rvert \ll 1$, from time $\tau_*$ to the end of the inflationary epoch. We then observe that geometric particle densities are typically larger for modes which leave the Hubble horizon earlier.

These outcomes are similarly found within the gravitational particle production mechanism, where it has been shown that super-Hubble modes dominate over sub-Hubble ones \cite{lyth}. 

Hence, we focus on particle production from time $\tau^*$, corresponding to the horizon crossing of the pivot scale $k^*$, to inflationary graceful exit, so to consider the simplified solutions, Eqs. \eqref{subfluc}, \eqref{supfluc}, for the fluctuations of the field, respectively, with 
\begin{equation}
    \label{setmom}
\{ {\bf k}_i \}= \begin{cases}  a(\tau_i)H_I < \lvert {\bf k}_1 \rvert < a(\tau_*) H_I, \\ a(\tau_*) H_I < \lvert {\bf k}_2 \rvert < a(\tau) M_{Pl},   \end{cases}
\end{equation}
where, on one hand, we impose as UV cutoff for comoving momenta the Planck mass, and on the other hand, the IR cutoff discards all modes already outside the Hubble horizon at the onset of inflation \cite{belfiglio2023superhorizon}.
Within the aforementioned choice, we are then neglecting the contribution of all modes that cross the Hubble horizon after $\tau_*$. In this regard, the total number of particles computed in this work necessarily represents an underestimate.

Further, we compute quartet probabilities from pairs with the same momentum $q_1$ and $q_2$, thus preserving the superhorizon approach introduced in Eq. \eqref{setmom}. Our results are shown in Figs. \ref{fig part pairs} and \ref{fig part quartets}.

Before moving to the Einstein frame, we underline again that the prescription for particle production is affected by a fundamental issue, namely the choice of the vacuum. For this reason, we mainly focus on comparison between different production mechanisms here, instead of attempting to derive total number densities. As previously discussed, the most widely used vacuum in quantum field theory in curved spacetime is the Bunch-Davies vacuum, which, however, suffers from infrared divergence. Thus, when dealing with the inflationary scenario, the magnitude of the field fluctuations, along with the Bogoljubov coefficients, is overestimated, leading to unrealistic results. This issue should be avoided when dealing with number density ratios, thus canceling the systematic error introduced.


\subsection{Does particle production change indicating a \emph{frame problem}?} \label{sec 4.3}

Particle production has been extensively explored in the Jordan frame. Now, we proceed to characterize these mechanisms in the Einstein frame to provide further evidence regarding the frame equivalence problem.

In this context, we begin with the transformed action reported in Eq. \eqref{ein action}, and we proceed in close analogy to the Jordan framework in deriving the corresponding equations. The inflaton field is split into a background component and a fluctuation term
\begin{equation}
    h({\bf x},t)=h_0(t)+\delta h({\bf x},t)\,, \label{field ein} 
\end{equation}
while the metric includes the conformal factor $\Omega(h)$
\begin{equation}\label{perturbed metric ein}
    g_{\mu\nu}^E=\left(g_{\mu\nu}^{(0)}+\delta g_{\mu\nu}\right)\Omega^2(h)=a(\tau)^2\Omega^2(h)\left(\eta_{\mu\nu}+2{ \Psi}\delta_{\mu\nu}\right)\,.
\end{equation}
Within this framework, the dynamical evolution of the field $h$ is governed by the Klein-Gordon equation
\begin{equation}\label{kg ein}
    \frac{1}{\sqrt{-g^E}}\partial_\mu\left(\sqrt{-g^E}g^{\mu\nu}_{E}\partial_{\nu}h\right)=\frac{\partial V_E(h)}{\partial h}\,,
\end{equation}
and, to impose canonical quantization, we express the field fluctuations in Fourier space as
\begin{equation}\label{fourierquant ein}
    \hat{\delta h}=\frac{1}{(2\pi)^{3/2}}\int{d^3k\left(\hat{a}_k\delta h_k(\tau)e^{i{\bf k\cdot x}}+\hat{a}_k^\dagger\delta h_k^*(\tau)e^{-i{\bf k\cdot x}}\right)}\,.
\end{equation}
The general expression for the field Fourier components is given by
\begin{equation}\label{fluctmotion1EIN}
    \delta h_k^{\prime \prime}+2 \mathcal{H} \delta h_k^\prime+\left[ k^2 + V^{E}_{, h h}a^2\Omega^2 +6 \epsilon \mathcal{H}^2 \right] \delta h_k = 0\,,
\end{equation}
which, by exploiting the usual rescaling $\delta\gamma_k\,=\,a(\tau)\delta h_k$, can be rewritten as
\begin{equation}\label{lle}
    \delta \gamma_k^{\prime \prime}+ \left[ k^2-\frac{1}{\tau^2} \left( (2+3\epsilon)+6\epsilon- \frac{ V^{E}_{, h h}}{H_I^2}\Omega^2 \right)    \right] \delta \gamma_k=0\,.
\end{equation}
In this way, we obtain an equation with the same functional form of Eq. \eqref{jor dyn pp}. Consequently, the solutions can be expressed again in terms of Hankel functions, as in Eq. \eqref{generic sol}, by replacing the index $\nu$ with
\begin{equation}\label{nuE}
    \nu_E\equiv \sqrt{\frac{1}{4}+(2+3\epsilon)+6\epsilon-V^{E}_{,h h}\Omega^2/H_I^2}\,.
\end{equation}
Hence, during the inflationary epoch, the field evolves in terms of the Hankel functions in both frames.

The geometric particle production mechanism in the Einstein frame is constructed by starting from the corresponding energy-momentum tensor
\begin{equation}
    T_{\mu \nu}^{(0),E}= \partial_{\mu} \delta h\  \partial_{\nu}\delta h- g_{\mu \nu}^{(0),E} \left[ \frac{1}{2}g^{\rho \sigma}_{(0),E}\  \partial_{\rho} \delta h\  \partial_{\sigma} \delta h - V_E(\delta h)  \right]\,,
\end{equation}
and the interaction Lagrangian
\begin{equation}
    \mathcal{L}_{I}^E=-\frac{1}{2}\sqrt{-g^{(0),E}}\delta g^{\mu\nu}_{E}T^{\left(0\right),E}_{\mu\nu}\,.
\end{equation}
Finally the probability amplitudes for producing particle pairs and quartets can be computed similarly to the Jordan frame, but considering
\begin{align} \label{Ampl e}
A_0^E({\bf x},\tau)=&  \Psi\Omega^2 e^{-i({\bf p_1}+{\bf p_2})\cdot {\bf x}}\bigg[\partial_0 \delta h_{p_1}^*\  \partial_0 \delta h_{p_2}^* \notag \\
&- \big( \frac{1}{2}\eta^{\rho \sigma} \partial_\rho \delta h_{p_1}^* \ \partial_\sigma \delta h_{p_2}^*-a^2 \Omega^2\tilde V^E_2(\delta h) \big)\bigg]\,, \notag\\ \\
 A_i^E({\bf x},\tau)=& \Psi\Omega^2e^{-i({\bf p_1}+{\bf p_2})\cdot {\bf x}} \bigg[\partial_i \delta h_{p_1}^*\  \partial_i \delta h_{p_2}^*\notag \\
    & + \big( \frac{1}{2}\eta^{\rho \sigma} \partial_\rho \delta h_{p_1}^* \ \partial_\sigma \delta h_{p_2}^*-a^2 \Omega^2\tilde V_2^E(\delta h) \big)\bigg] ,\notag\\ \\
    B_0^E({\bf x},\tau)=&-B_i^E({\bf x},\tau)= \notag\\
=&\Psi a^2\Omega^4 \tilde V_4^E(\delta h)e^{-i({\bf q_1}+{\bf q_2}+{\bf q_3}+{\bf q_4})\cdot {\bf x}}\,.
\end{align}

In order to extend gravitational particle production to the Einstein frame, we need to include the transition to the reheating stage. From Eq. \eqref{fluctmotion1EIN}, we get the corresponding equation for reheating, modeled as before with a scale factor defined as in Eq. \eqref{reheating}
\begin{equation}\label{fluctmotion1EINreheating}
   \delta \gamma_k^{\prime \prime}+ \left( k^2-2\frac{1+12\epsilon}{\tau^2} +\Omega^2\frac{H_{re}^4\tau^4}{16} \right)     \delta \gamma_k=0\,.
\end{equation}
We expect the field to undergo an adiabatic evolution, leading to the same solution obtained for the Jordan frame, namely Eq. \eqref{adiab eq}, with 
\begin{equation}
    \omega_E^2=k^2-2\frac{1+12\epsilon}{\tau^2} +\Omega^2\frac{H_{re}^4\tau^4}{16}\,.
\end{equation}

Within this picture, recalling the definition of Bogoljubov coefficients in Eqs. \eqref{alpha}--\eqref{beta}, we compute particle production through both geometric and gravitational mechanisms. Specifically, Tab. \ref{fig confronto} shows the averaged ratio between particle density in the Einstein and in the Jordan frame. 
We do not consider the entire evolution of these ratios because they are nearly constant, with a maximum semi-dispersion of less than $1\%$.
The resulting comparison reveals a disparity between the two regimes, and in particular, each potential presents a proper, nearly-constant ratio which consistently differs from 1. 
This suggests a strong dependence on the energy scales involved and on the behavior of the single potentials.

\begin{table}[ht]
  \centering
      \begin{tabular}{c|c|c|c}
    \hline\hline
     & $|\mathbf{\mathcal{C}_{{\bf p}_1,{\bf p}_2}^E|^2/|\mathcal{C}_{{\bf p}_1,{\bf p}_2}^J|^2}$& $\mathbf{|\mathcal{C}_{{\bf q}_1^2,{\bf q}_2^2}^E|^2/|\mathcal{C}_{{\bf q}_1^2,{\bf q}_2^2}^J|^2}$ & $\mathbf{|\beta_k^E|^2/|\beta_k^J|^2}$ \\
    \hline
    \hline
    $V_S$ & $0.47$& $0.94$ & $0.91$   \\
    
    $V_E$ & $0.01$& $0.03$ & $1.34$   \\
    
    $V_T$ & $0.08$& $0.00$ & $1.11$  \\
    
    $V_N$ & $0.07$& $1.07$    & $1.02$\\  
     \hline\hline\end{tabular}
  \caption{Comparison between particle production, respectively in the Jordan and Einstein frames, fuelled by inflationary potentials. Specifically: pair production densities with $p_2=1000/\tau_f$, averaged over $p_1\in[-1/\tau_0,-10/\tau_0]$; quartet production densities with $q_2=1/\tau_f$, averaged over $q_1\in[1/\tau_f,10/\tau_f]$; gravitational particles density, represented by the Bogoljubov coefficient $|\beta_k|^2$, averaged over $k\in[-1/\tau_0,-10/\tau_0]$.}
  \label{fig confronto}
\end{table}

\begin{figure*}[ht]
  \centering
  \textbf{a)}\hspace{0.mm}
\includegraphics[width=.44\linewidth]{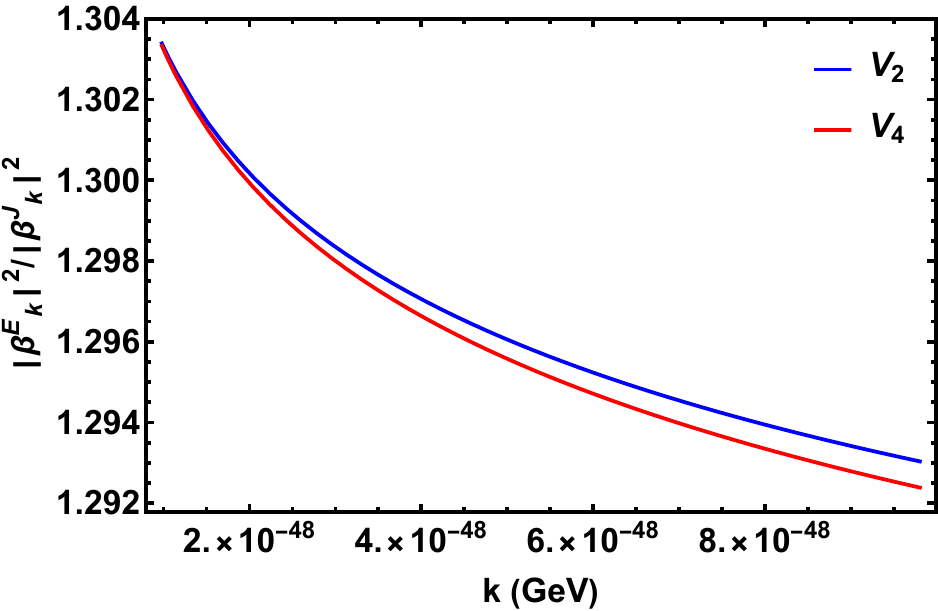}
  \hspace{4mm}
  \textbf{b)}\hspace{0mm}
\includegraphics[width=.46\linewidth]{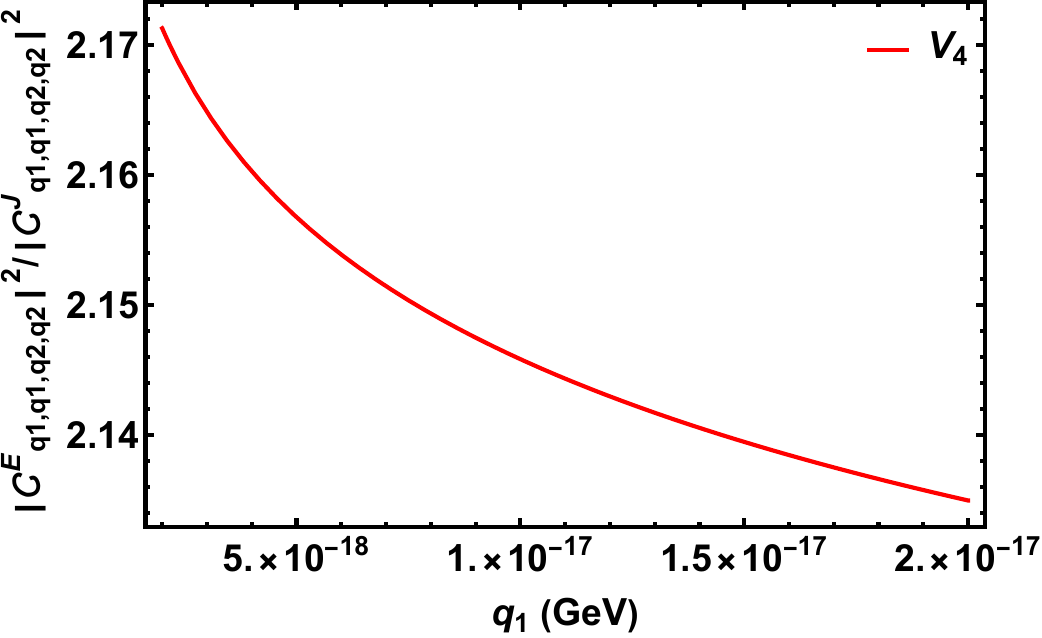}
  \hspace{0.mm}\vspace{3mm}

   \textbf{c)}\hspace{0mm}  
    \includegraphics[width=.46\linewidth]{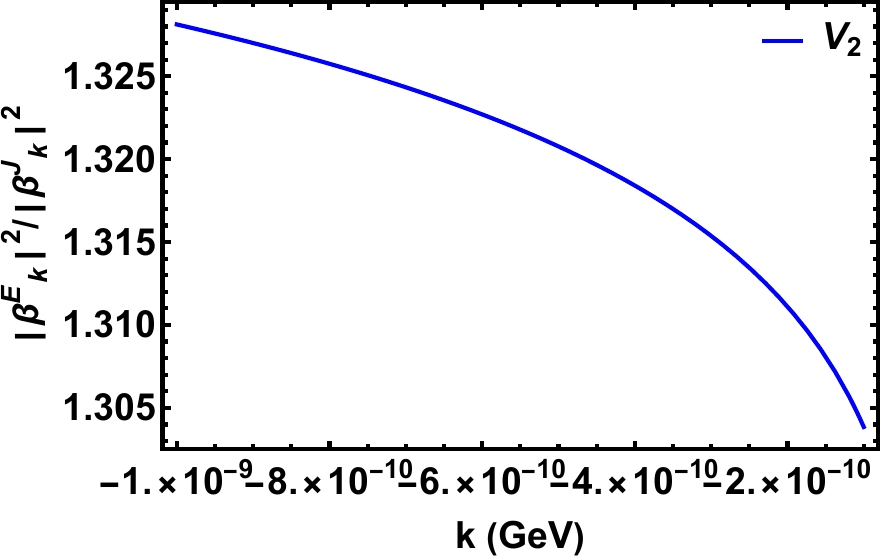}\hspace{1mm} 
  \caption{Comparison between particle production in the Jordan and Einstein frames, respectively. Specifically: {\bf a)} pair production densities during inflation, governed by the potentials $V_2$ and $V_4$ with $p_2=1000/\tau_f$; {\bf b)} gravitational particle density, represented by the Bogoljubov coefficient $|\beta_k|^2$, during the inflation-to-reheating transition, governed by the potentials $V_2$ and $V_4$ with $q_2=1/\tau_f$; {\bf c)} quartets production densities during inflation, governed by the potentials $V_4$ with $q_2=1/\tau_f$; {\bf d)} gravitational particle density during a matter-to-radiation-like transition, governed by the potential $V_2$ with $q_2=1/\tau_f$.}
  \label{fig frame issue}
\end{figure*}

Our findings show that geometric particle production is significantly affected by the frame choice. 
This phenomenon can be traced back to the presence of the non-minimal Yukawa-like coupling within the inflationary potential, that inevitably affects perturbative probability amplitudes.
These outcomes then provide a hint of the quantum non-equivalence between the two frames.
Indeed, classically, the inflationary dynamics is unaltered under frame transformation, while this symmetry is broken when second quantization of the inflaton comes into play.
Specifically, looking at Eq. \eqref{conf transf}, it can be noticed that, when transforming the Ricci scalar, the inflaton field is involved. Although this does not cause issues classically, when we quantize the transformed field $h$, we discard part of the energy of the inflaton itself that is encoded in the scalar curvature.
To further address such a quantum frame non-equivalence, in the next subsection we provide some additional remarks concerning particle production mechanisms in the Einstein frame.


\subsection{Revising the frame problem}

As mentioned above, our outcomes for inflationary particle production in Jordan and Einstein frames lead to a quantum frame problem. 
In this respect, we wonder how this phenomenon manifests in different cosmological scenarios.
First of all, as previously discussed, we remark that strong coupling constants cannot be considered within this framework, as they result in unphysical outcomes.
Thus, we start from a generic scalar field, specifically fueled by quadratic and quartic power-law potentials, and we test it in inflation as well as in other cosmological epochs.
The aforementioned potentials can be written, {\bf in their simplest form}, as
\begin{align}
    V_2(\phi)&=\frac{1}{2}m^2\phi^2\,,\\
    V_4(\phi)&=\frac{1}{4}\lambda\phi^4\,,
\end{align}
where the free parameters are fixed as $m=10^{11}$ GeV and $\lambda=2\cdot10^{-17}$. To address power-law models, we first impose the same energy bounds as in Sect. \ref{sect 4}, and then we recover all the necessary quantities following the previous prescriptions.
On one hand, regarding geometric pair production the ratios of the averaged densities are given by
\begin{equation}
    R_2=0.36\,,\quad\,R_4=0.67.
\end{equation}
On the other hand, as shown in Fig. \ref{fig frame issue}, gravitational and geometric quartet productions yeld comparable values with respect to the previously addressed models.

By investigating further transitions, such as the radiation-to-matter or matter-to-vacuum energy transitions, the field dynamics obeys to Eqs. \eqref{ll} and \eqref{lle}, respectively in the Jordan and the Einstein frame, with $\omega^2$  determined by the choice of the corresponding cosmological scale factor, as given by Eqs. \eqref{desitter}, \eqref{reheating}, \eqref{radiation}.
Specifically, the energy of the scalar field, and consequently the interaction, is too low to generate a significant difference in particle production between the two frames.
This outcome validates our thesis regarding the Ricci scalar, \textit{i.e.}, at small energy scale for the field, for the interaction term, and $\Omega(\phi)$ , the contribution to the total particle production is negligible.

Hence, to ensure the generality of our statement, we introduce a toy model where the non-minimally coupled scalar field is sub-dominant during the transition from a matter to a radiation phase, in close analogy to a transition between a matter-like reheating and the radiation-dominated era, but with energy scales comparable to inflation.
During the matter-dominated era, the field evolves approximately as in Eq. \eqref{gen sol field} with 
\begin{equation}
    \nu\simeq \left[\frac{1}{4}+2(1-6\xi)\right]\,,\quad\nu_E\simeq\frac{3}{2}\,,
\end{equation}
while in the radiative era, we expect an adiabatic phase with
\begin{equation}
    \omega= k^2+V^{\prime\prime}H_r^2\tau^2.
\end{equation}
In Fig. \ref{fig frame issue}, we also report the ratio between the particle densities obtained in the Jordan and Einstein frames, respectively. 
We remark that, during this matter-to-radiation transition, we restrict our analysis to gravitational particle production, due to the complexity of extending the geometric mechanism to this context, particularly when dealing with the dynamics of spacetime perturbations after the inflationary epoch. Moreover, quantum fluctuations associated with the scalar field are typically smaller in this case, implying that geometric particle production from such fluctuations is significantly reduced.

Once again, the number of particles produced differs between the two frames.
In other words, we can safely state that the quantum frame non-equivalence is a phenomenon not limited to inflation, but is a general issue affecting all epochs, particle regimes, and models. 
In particular, anytime the energy is sufficient, such a non-equivalence becomes evident, thus clearly demonstrating a quantum frame issue related to the role of the scalar field in the conformal transformation between the frames. 
Hence, by quantizing the scalar field, we miss part of the information embedded inside the Ricci scalar, that acts as a classical field.
Within this picture, the two frames can not be equivalent unless a quantum theory of gravity may quantize the scalar curvature and thus potentially prove equivalent behavior in the Jordan and Einstein frames.


\section{Theoretical consequences}\label{sect 6}

In this work, we delve into extensions of well-fitting single-field inflationary models to the presence of non-minimal field-curvature coupling \cite{planck}. 
We characterize the inflationary dynamics by considering different potentials driving the scalar inflaton field, focusing on particle creation processes arising from spacetime expansion and quantum fluctuations and pointing out an underlying quantum non-equivalence between the Jordan and Einstein frames. We here discuss step by step the outcomes of our work. 

\subsection{Non-minimally coupled inflation}

After studying several models, we conclude that the inflationary epoch is consistent with non-minimal coupling, especially for small interaction strengths, both positive and negative.  
The choice of weak interactions, analogously to particle physics, is necessary for the inflaton to undergo a slow-roll phase, as shown in Sect. \ref{sect 3.1.2}, where we explicitly rule out the conformal coupling prescription. In addition, strong interactions would lead to a non-physical effective gravitational constant, as shown by Eq. \eqref{geff}. It is worthy to notice that this achievement is general and holds independently of the inflationary potential under consideration.

Large field models are preferred over small field ones, providing further confirmation for Starobinsky-like models. In particular, the hilltop and the quartic $\alpha-$attractor models do not present a suitable behavior when coupled with the scalar curvature. Lastly, we highlight that, when working in a weakly interacting regime, the choice between positive and negative coupling does not significantly affect the dynamics; hence, for the sake of brevity, we proceed considering only positive coupling constants.

\subsection{Gravitational and geometric particle production}

As discussed in the previous section,  the main problem associated with a proper evaluation of particle densities in curved backgrounds is represented by the choice of the initial vacuum state for the field.

Specifically, for inflationary fluctuations, we adopt the Bunch-Davies vacuum, representing the most used  \emph{in-}vacuum choice currently available \cite{Bunch:1978yq} and being a local attractor in the space of initial states for an expanding background.

In addition, we characterize the reheating era to analyze how inflaton fluctuations manage to provide a well-defined \emph{out}-vacuum state after inflation. We show that the quantized fluctuations satisfy the conditions of adiabatic expansion, Eq. \eqref{adiab}, during the reheating stage, so that the cosmological background is complete and consistent to properly address particle production mechanisms.

Due to the infrared divergence, we notice that inflaton fluctuations, and consequently particle densities, \emph{increase non-physically} in the limit of zero momentum. 

To address this, we mainly focus on momentum ranges far from the divergence, introducing a proper infrared cutoff which neglects modes outside the Hubble radius at the beginning of inflation. Accordingly, we  consider ratios between geometric and gravitational number densities within such ranges.

Nevertheless, we clearly highlight that this issue is still present and thus requires further investigations to be refine the vacuum choice at the inflationary onset. In principle, regularization or renormalization techniques should be taken into consideration to heal such divergences. 

Alternatively, we may need to identify a more appropriate vacuum instead of the conventional Bunch-Davies vacuum.

In this scenario, we explore particle production through gravitational and geometric mechanisms. To determine the particles number densities, we start from a perturbative approach, approximating the inflationary potentials as  series of even powers up to fourth order. We focus on production during the inflationary regime and then we consider the transition to the reheating stage, to properly address the inflaton dynamics when the slow-roll approximation is no longer valid. Further, the total number densities arising from gravitational production has been shown to be zero within a pure de Sitter-like evolution.

In computing the total amount of particles produced during inflation and reheating, 

\begin{itemize}
    \item[-] we underestimate the geometric particle production of pairs, as due to numerical complexity, 
    \item[-] we find that this contribution is almost the same as the gravitational one, 
    \item[-] however, we notice that the production of quartets still remains subdominant.
\end{itemize}

Summarizing, we can safely conclude that \emph{the geometric particle production significantly contributes to the total amount of particles produced during inflation and reheating}. 

In particular, if dark matter creation can be traced back to such mechanisms, future research should explore both geometric and gravitational scenarios.

\subsection{Addressing the frame problem through particle production}

The analysis of the inflationary dynamics involves addressing the frame issue. We recover the inflationary evolution in both the Jordan and Einstein frames, obtaining

\begin{itemize}
    \item[-] dynamically the results turn out to be perfectly consistent, 
    \item[-] accordingly, the above outcome supports the \emph{equivalence between the two frames at a classical level}.
\end{itemize} 
However, when investigating particle production in both frames, thus introducing second quantization of the inflaton field, we find that, from such quantum perspective, the two frames leads to different particle amounts.
Further, when generalizing this picture to other cosmological eras, we can safely assert that this phenomenon is independent of the background evolution.
We underline that, at low energy scales, the difference between the frames is typically negligible, but considering energy scales comparable to those of inflation, such departures are evident when computing particle densities.
The disparity in particle production arises from the conformal transformation which defines the Einstein frame and thus makes the Ricci curvature dependent on the scalar field, so that a classical quantity now encodes a quantum field.

Consequently, when the system is quantized, part of the information is locked into the Ricci scalar, preventing it from contributing to particle production.

This relationship, along with the energy scale dependence, provides further evidence that a full quantum theory of gravity is ultimately required to properly address the issue of frame equivalence.


\section{Final outlooks and perspectives}\label{sect 7}

In this paper, we compared two mechanisms of cosmological particle production associated with primordial fluctuations, assuming the scalar inflaton field to be non-minimally coupled to the spacetime curvature through a Yukawa-like interacting Lagrangian. 

More precisely, we focused on gravitational and geometric particle production and, in particular, we studied under which conditions particles obtained from the geometric term entering the interacting Lagrangian, namely when the Bogoljubov coefficients are set to zero, can significantly contribute to the overall production. 

In other words, despite the geometric particle production is obtained up to  second order of our perturbative approach, it is possible to demand for a net contribution from these particles that cannot be neglected  throughout the entire inflationary phase, resulting into a modification of the net energy momentum tensor.

To explore this possibility, different classes of inflationary models have been investigated, focusing on small, large and natural inflationary paradigms. 

To do so, we first analyzed the dynamics of non-minimally coupled inflationary models and concluded that:\\ 

\begin{itemize}
\item[-] small field inflation is severely disfavored when the inflaton field is non-minimally coupled to curvature. In this respect, the hilltop potentials are ruled out, while natural inflation remains dynamically acceptable, 
\item[-] the here-involved large field potentials namely the Starobinsky and $\alpha$ attractor models, fit much better, albeit with some inconsistencies in the case of $\alpha$ attractor potentials, where the chaotic behavior can be achieved in quadratic cases only, suggesting that  \emph{Starobinsky-like} models remain extremely favored in describing the inflationary phase. 
\end{itemize} 

Our dynamical analysis proved that the strength of the non-minimal couplings, across  the inflationary regime, might be carefully fine-tuned. Indeed, regardless  the potential under investigation, only weak couplings yield consistent slow-roll inflation, ruling out the widely-consolidate conformal coupling.

Afterwards, by introducing the unavoidable quantum fluctuations associated with the inflaton field, and the corresponding spacetime inhomogeneities generated by such fluctuations, we determined their dynamics from an initial Bunch-Davies vacuum state up to the reheating phase.
In particular, we verified that, before vanishing, the fluctuations reach a state of adiabatic expansion, so that their particle excitations can be properly defined during reheating.

Bearing this in mind and, in particular, defining  initial and final states allowed us to investigate particle production across the different models we introduced. Specifically, exploring  perturbatively the aforementioned gravitational and geometric mechanisms, namely introducing the S-matrix formalism and approximating the inflationary potentials through power series, we observed that the net number of particles produced in the Jordan and Einstein frames differs significantly. 

Phrasing it differently, while dynamically speaking there exists a net equivalence between the Jordan and Einstein frames, as proved for small coupling constants, the number of particles appears quite different passing from a frame representation to another one, providing a strong frame problem consisting in the discrepancy between the two frames, inevitably caused by the presence of non-minimal coupling. 

As a possible explanation, the non-equivalence between frames can be  attributed to the conformal transformation between the frames themselves, and, in principle, this issue may only be healed within a fully-consistent, but so far missing, quantum gravity puzzle.

Further, as widely discussed throughout the paper, we noted that particle production in curved spacetime is generally affected by the issue of vacuum choice and, accordingly, the presence of infrared divergences when starting from a Bunch-Davies vacuum state. 
In this respect, we proceeded by comparing the geometric and gravitational mechanisms instead of providing total number densities in both frameworks, in the attempt to alleviate this sort of systematic discrepancy.

In future works, we aim to better understand the physics related to these fundamental issues by exploring alternative vacuum states. Further, we intend to employ renormalization techniques, to check whether the same issue can be faced using the renormalization group.

Additionally, our purpose is to propose a new particle production mechanism involving inhomogeneities beyond the perturbative approach, developing a path integral strategy that could also improve consistency with non-polynomial models.

Finally, to give a more complete and comprehensive picture of the inflationary phase, we plan to include possible additional couplings, addressing for example the presence of electromagnetic fields.

Indeed, the presence of geometric particles can be associated with the existence of exotic constituents, that may contribute to the net energy-momentum budget of the universe under certain assumptions. This class of particles can be reinterpreted as quasi-particles of geometry, as found in Refs. \cite{Belfiglio:2022cnd,Belfiglio:2022yvs,Belfiglio:2023eqi,Belfiglio:2023rxb,belfiglio2023superhorizon,Belfiglio:2024xqt} and, so, more efforts in this regards will shed light on the physics of such particles whose origin comes from inflationary domain.


\section*{Acknowledgements}
AB is grateful to Fabio Costa for engaging discussions on subjects related to this work. OL sincerely expresses his thankfulness to Roberto Della Ceca, Roberto Franzosi, Luigi Guzzo, Stefano Mancini and Marco Muccino for support and discussions. He is also in debit with Eoin 
\'O Colg\'ain and Mohammad Mehdi Sheikh-Jabbari for private discussions on topics connected to this work. TM acknowledges the Brera National Institute of Astrophysics (INAF) for financial support and Hernando Quevedo for interesting discussion on topics related to fields and, in particular, on the great importance of non-minimal couplings in inflationary stages and, in general, at primordial times. 

%


\appendix

\section{Characterizing the Jordan and the Einstein frames }\label{appendix1}

In the Jordan frame, we proceed as usual by varying the total action including field-curvature coupling, obtained from Eq. \eqref{L tot}, to derive the modified Einstein field equations. Hence, we get the standard Hilbert-Einstein Lagrangian variation along with an additional term resulting from the variation of the non-minimal coupling given by Eq. \eqref{L tot} 
\begin{equation}
    \delta\left(\xi R\phi^2\sqrt{-g}\right)=\xi\delta R_{\mu\nu}g^{\mu\nu}\phi^2\sqrt{-g}+\xi G_{\mu\nu}\phi^2\sqrt{-g}\delta g^{\mu\nu}\,.
\end{equation}
There, we used the definition of the Einstein tensor $G_{\mu\nu}\equiv R_{\mu\nu}-\frac{1}{2}g_{\mu\nu}R$. Further, it can be shown that
\begin{equation}
   g^{\mu\nu}\delta R_{\mu\nu}=\left(g_{\mu\nu}\nabla_\alpha\nabla^\alpha-\nabla_\mu\nabla_\nu\right)\delta g^{\mu\nu}\,,
\end{equation}
where $\nabla_\mu$ represents the covariant derivative.
The energy-momentum tensor is then derived as
\begin{equation}
    T^{int}_{\mu\nu}=\frac{2}{\sqrt{-g}}\frac{\delta S_{int}}{\delta g^{\mu\nu}}=\xi\left(g_{\mu\nu}\nabla_\alpha\nabla^\alpha-\nabla_\mu\nabla_\nu\right)\phi^2+\xi\phi^2G_{\mu\nu}.
\end{equation}
Finally, the new Einstein equations read
\begin{equation}
\begin{split}
     &\frac{1-\chi\xi\phi^2}{\chi}G_{\mu\nu}=2\xi\phi\left(g_{\mu\nu}\nabla_\alpha\nabla^\alpha-\nabla_\mu\nabla_\nu\right)\phi+\\
    &\quad+\left(1-2\xi\right)\partial_\mu\phi\partial_\nu\phi-g_{\mu\nu}\left[\left(\frac{1}{2}-2\xi\right)\partial_\alpha\phi\partial^\alpha\phi-V(\phi)\right]. \label{efej}   
\end{split}
\end{equation}

On the other hand, to transform a non-minimally coupled Lagrangian in the Jordan frame into a minimally coupled model in the Einstein frame, we perform the conformal transformation, $g_{\mu\nu}^E=g_{\mu\nu}\Omega^2(\phi)\>\Rightarrow\> \sqrt{-g^E}=\sqrt{-g}\Omega^4(\phi)$ \cite{einsteinframe,einsteinframe1}, giving
\begin{equation}\label{conf transf}
R=\Omega^2(\phi)\left[R^E-6\left(\frac{\partial_{\mu}\Omega\partial^\mu\Omega}{\Omega^2(\phi)}+\nabla^\mu\nabla_\mu\ln{\Omega(\phi)}\right)\right],    
\end{equation}
where the label ``E" refers to the Einstein frame. The corresponding action for the field reads
\begin{equation}\label{ein action}
    S_E=\int{d^4x{\sqrt{-g^E}}\left[-\frac{R^E}{2\chi}\left(\frac{1-\xi\chi\phi^2}{\Omega^2}\right)+\mathcal{L}^E\right]},
\end{equation}
where $\mathcal{L}^E=\mathcal{L}\Omega^{-4}$, neglecting moreover the term $\nabla^\mu\nabla_\mu\ln\Omega$ as it does not affect the equations of motion.

The scale factor is determined so to recover the Hilbert-Einstein action, $S_{HE}=-\frac{1}{2\chi}\int{d^4x{\sqrt{-g}R}}$. Thus, we impose
\begin{equation}
    \Omega^2=1-\xi\chi\phi^2. \label{scale factor}
\end{equation}
In this way, we recover a new kinetic and potential term, respectively
\begin{subequations}
\begin{align}
        X_E&=\frac{1}{2}\partial^\mu\phi\partial_\mu\phi\left(\frac{\Omega^2+6\frac{1-\xi \chi\phi^2}{\chi}\left(\frac{\partial\Omega}{\partial\phi}\right)^2}{\Omega^4}\right),\\
        V_{E}&(\phi)=\frac{V(\phi)}{\Omega^4(\phi)}.\label{pot}
\end{align}
\end{subequations}

Within this transformed Lagrangian, we define a new field $h$ such that
\begin{subequations}    
\begin{align}
    &\frac{\partial h}{\partial \phi}=\frac{\sqrt{\Omega^2+6\frac{1-\xi\chi\phi^2}{\chi}\left(\frac{\partial\Omega}{\partial\phi}\right)^2}}{\Omega^2},\label{jacobian}\\
    &\frac{\partial\Omega}{\partial\phi}=-\frac{\xi \chi\phi}{\sqrt{1-\xi \chi\phi^2}},
\end{align}
\end{subequations}
whose explicit expression is obtained by combining Eqs. \eqref{scale factor} and \eqref{jacobian},
\begin{equation}\label{h}
\begin{split}
     h(\phi)&=\sqrt{\frac{6}{{\chi}}}\tanh^{-1}{\left(\frac{\sqrt{6\chi}\xi{\phi}}{\sqrt{1+\phi^2\chi\xi(-1+6\xi)}}\right)}\\
     &-\sqrt{\frac{{-1+6\xi}}{{\chi\xi}}}\sinh^{-1}{\left(\sqrt{\chi\xi(-1+6\xi)}\phi\right)}.
\end{split}
\end{equation}

\section{Plots }\label{appendix3}

\begin{figure*}[ht]

  \centering
  \textbf{a)}\hspace{0.mm}
    \includegraphics[width=.44\linewidth]{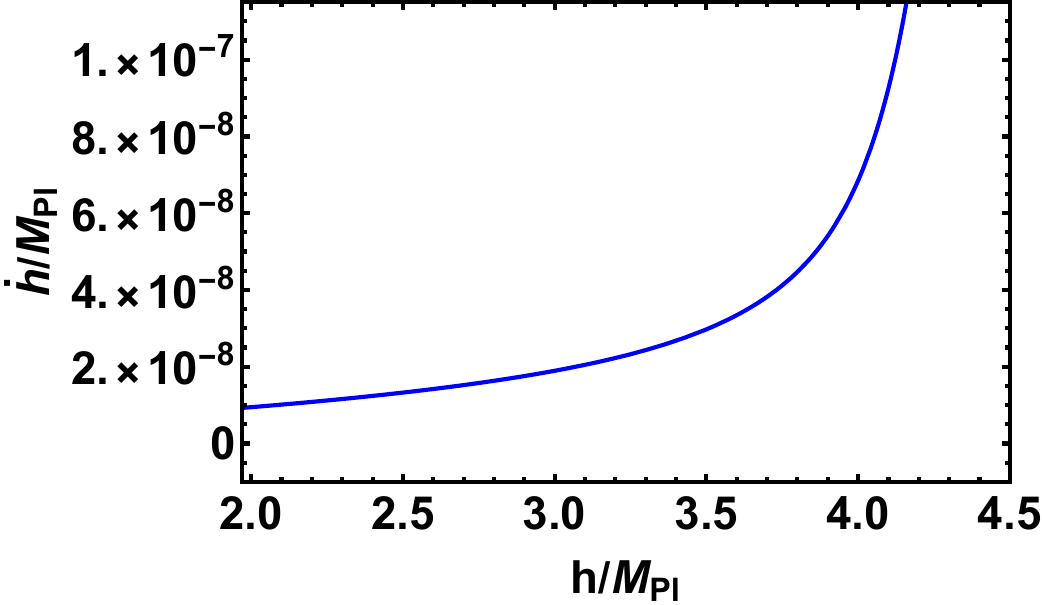}
  \hspace{0mm}
  \textbf{b)}\hspace{0mm}
    \includegraphics[width=.44\linewidth]{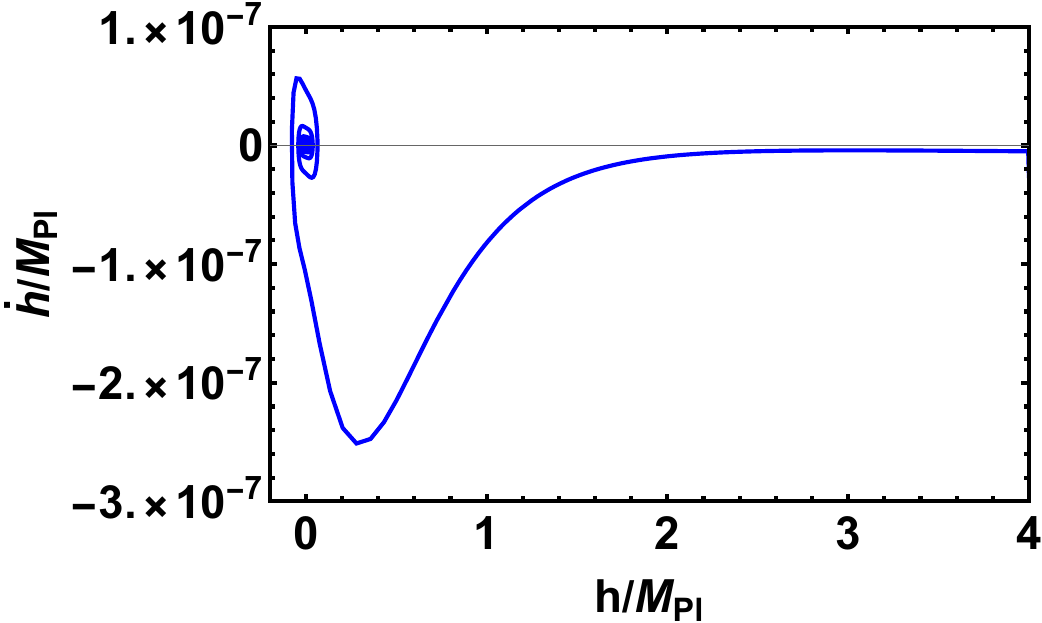}
  \caption{Configuration space of respectively: {\bf a)} the hilltop potential, Eq. \eqref{hill pot}, in the Eistein frame with $p=2$ and $\mu=4M_{Pl}$ \cite{planck}; {\bf b)} $E$-model, Eq. \eqref{E pot}, with $n=2$.}
  \label{fig not work}
\end{figure*}


In this appendix, we present plots related to the potentials introduced in Sect. \ref{Sect 2.3}. 

First, we explicitly show that, concerning the hilltop potentials and the quartic $\alpha$-attractor models, non-minimally coupled inflation is not viable. Indeed, looking at the dynamics in the configuration space, it is evident that there is not a well-defined attractor providing the graceful exit from inflation, as shown by Fig. \ref{fig not work}.
Specifically, in Fig. \ref{fig not work}, we only present the quadratic hilltop potential and the quartic $E$-model in the Einstein frame, as the other scenarios have similar features.

Then, we only consider models that provide a suitable inflationary scenario, such as the Starobinsky model, the $\alpha-$attractor with $n=m=1$, and natural inflation. 

\begin{figure*}[p]
  \centering
  \textbf{a)}\hspace{0.mm}
    \includegraphics[width=.41\linewidth]{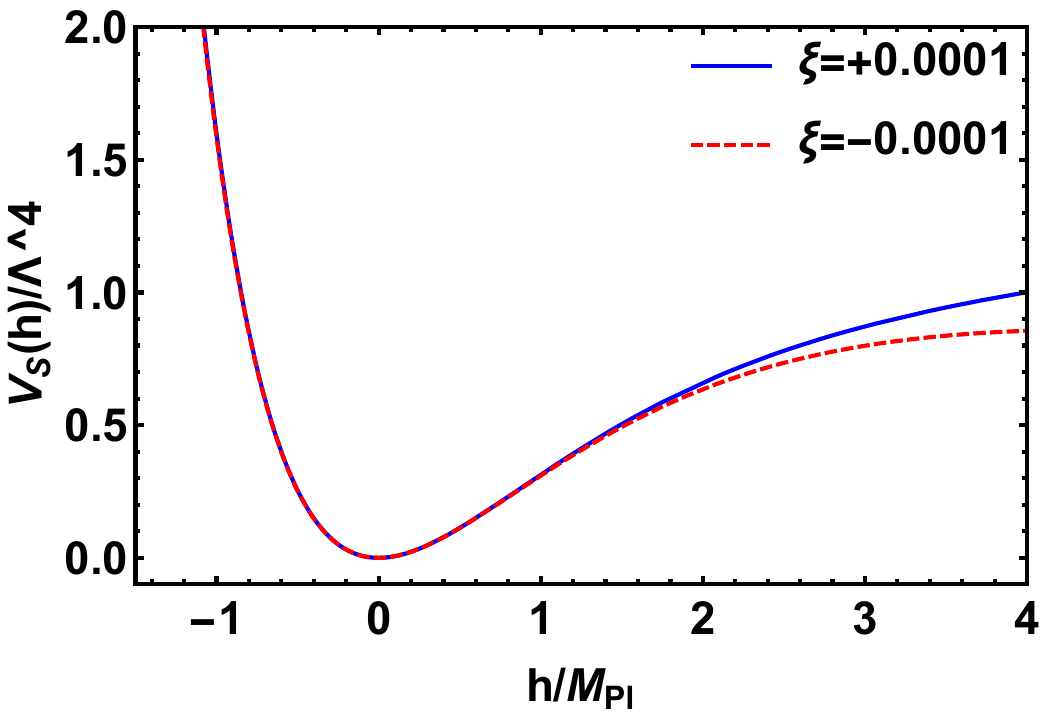}
  \hspace{0mm}
  \textbf{b)}\hspace{0mm}
    \includegraphics[width=.405\linewidth]{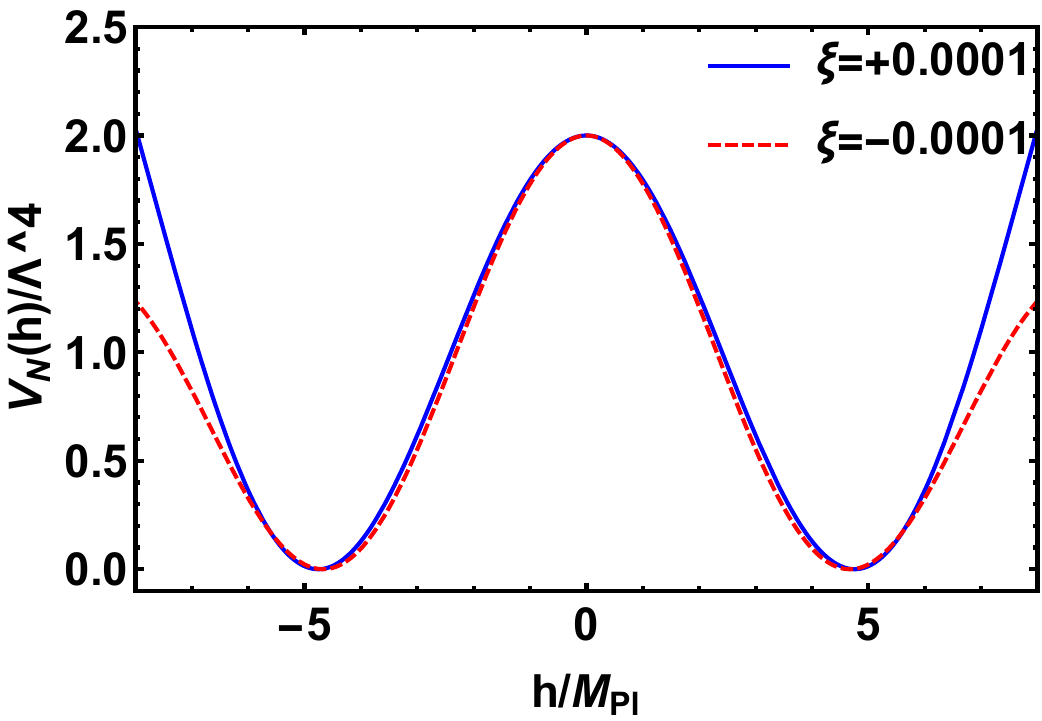}\\
    \vspace{0mm}
  \textbf{c)}
    \includegraphics[width=.41\linewidth]{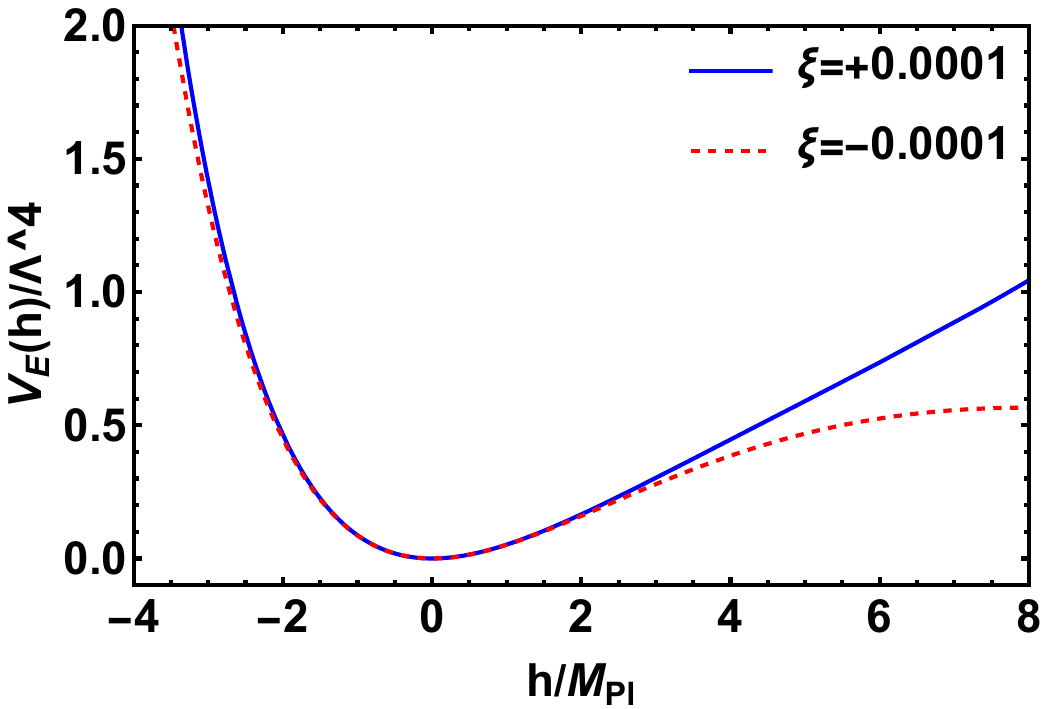}
  \hspace{0mm}
  \textbf{d)}
    \includegraphics[width=.41\linewidth]{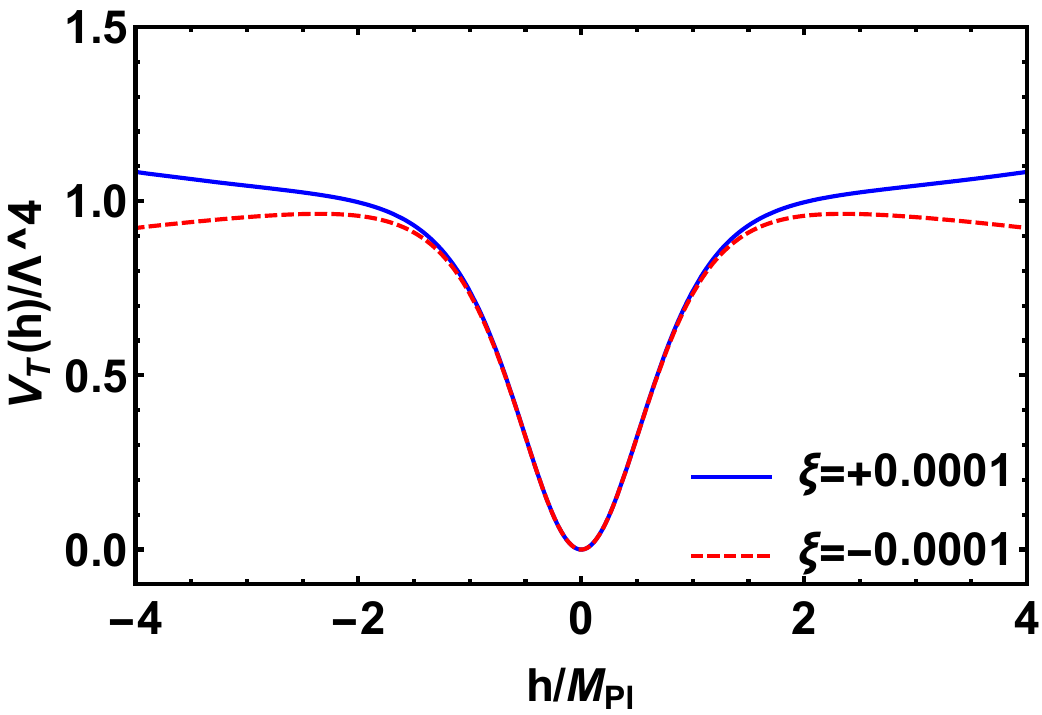}\\
     \vspace{0mm}
  \textbf{e)}\hspace{1mm}
    \includegraphics[width=.405\linewidth]{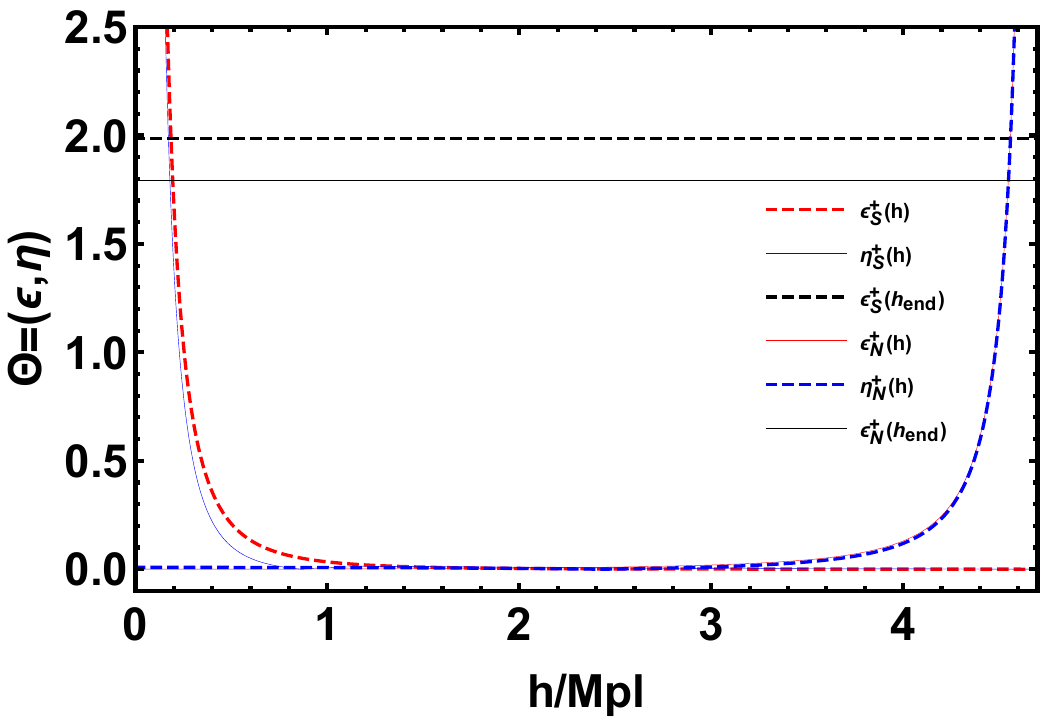}
  \hspace{0mm}
    \textbf{f)}\hspace{0mm}
    \includegraphics[width=.405\linewidth]{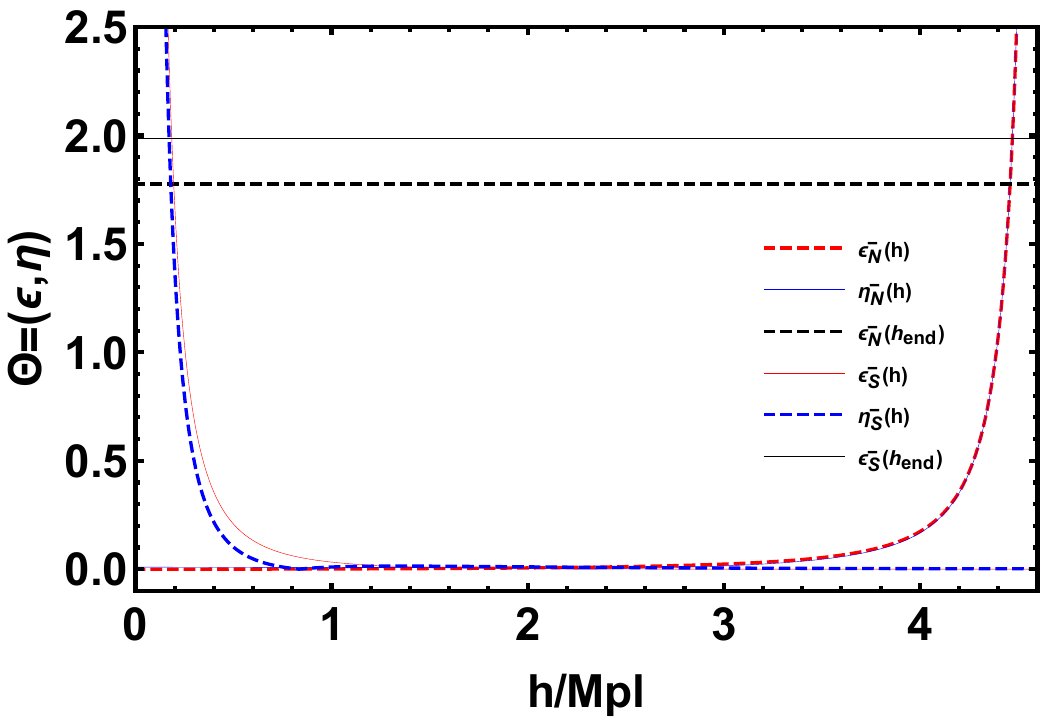}\\
    \vspace{0mm}
  \textbf{g)}\hspace{0mm}
    \includegraphics[width=.41\linewidth]{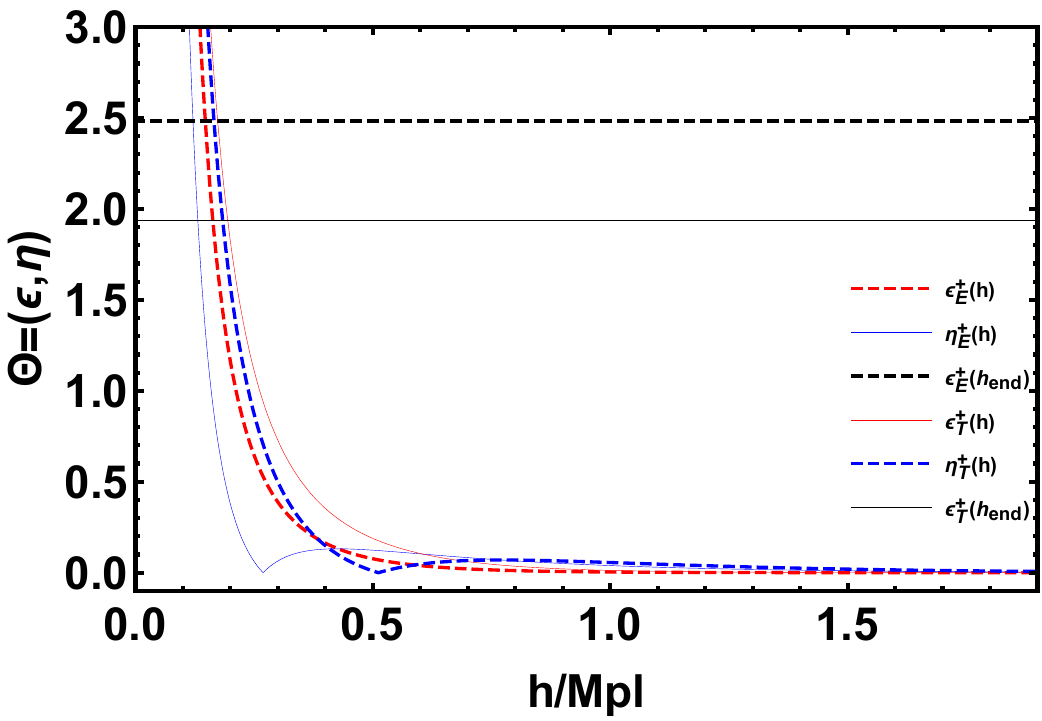}  \hspace{0mm}
    \textbf{h)}
    \includegraphics[width=.41\linewidth]{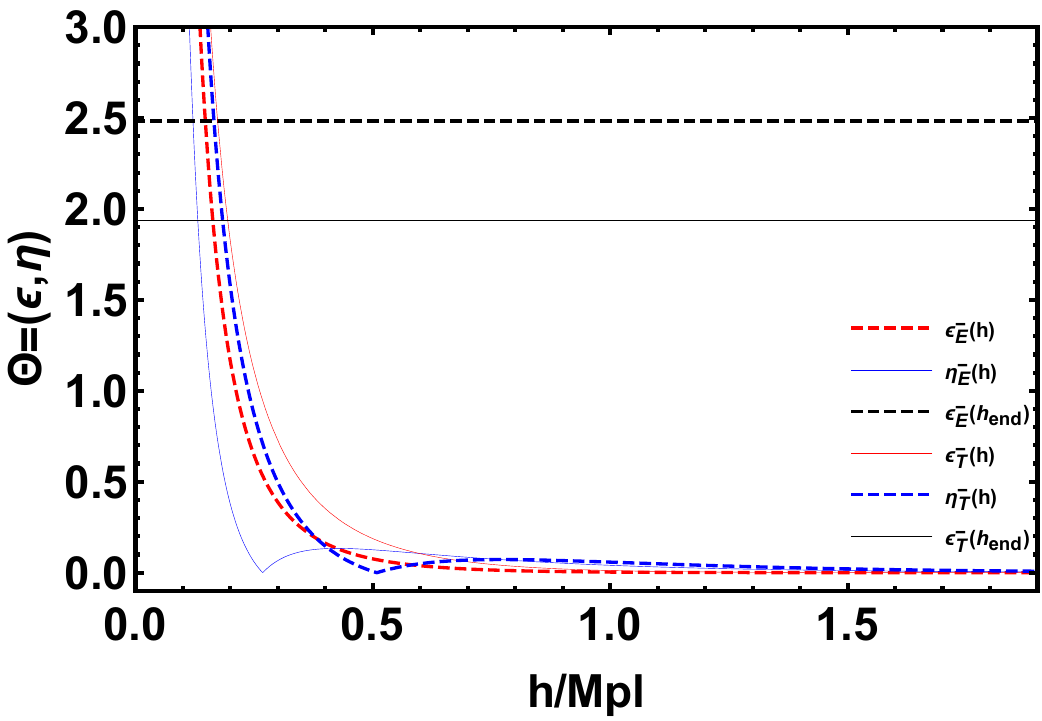}
  \caption{Comparison between positive and negative non-minimally coupled inflationary potentials introduced in Sect. \ref{Sect 2.3}: \textbf{a)}the Starobinsky potential $V^E_S(h)$, from Eq. \eqref{star}; \textbf{b)} the $E-$model $V^E_E(h)$, from Eq. \eqref{E pot}, with $\alpha^E=10$; \textbf{c)} the $T-$model $V^E_T(h)$, from Eq. \eqref{T pot}, with $\alpha^T=0.1$; \textbf{d)} natural inflation potential $V^E_N(h)$, from Eq. \eqref{nat pot}, with $f=1.5M_{Pl}$.
  Next, the corresponding slow-roll parameters: \textbf{e), g)} positive non-minimal coupling; \textbf{f), h)} negative non-minimal coupling.}
  \label{fig pot}
\end{figure*}

\begin{figure*}[p]
  \centering
  \textbf{a)}\hspace{0.mm}
    \includegraphics[width=.44\linewidth]{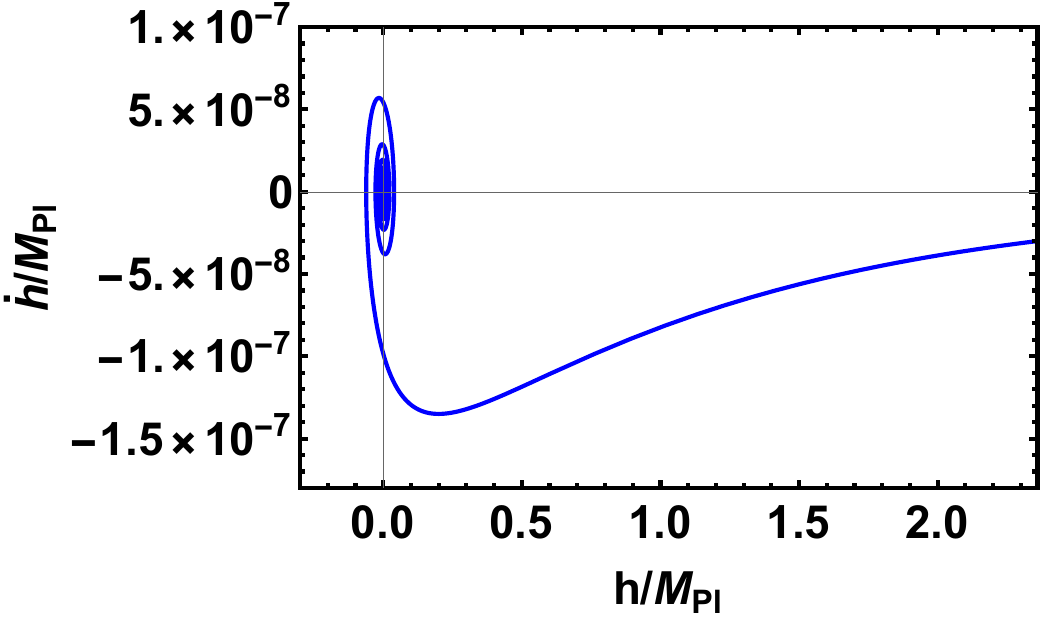}
  \hspace{0mm}
  \textbf{b)}\hspace{0mm}
    \includegraphics[width=.43\linewidth]{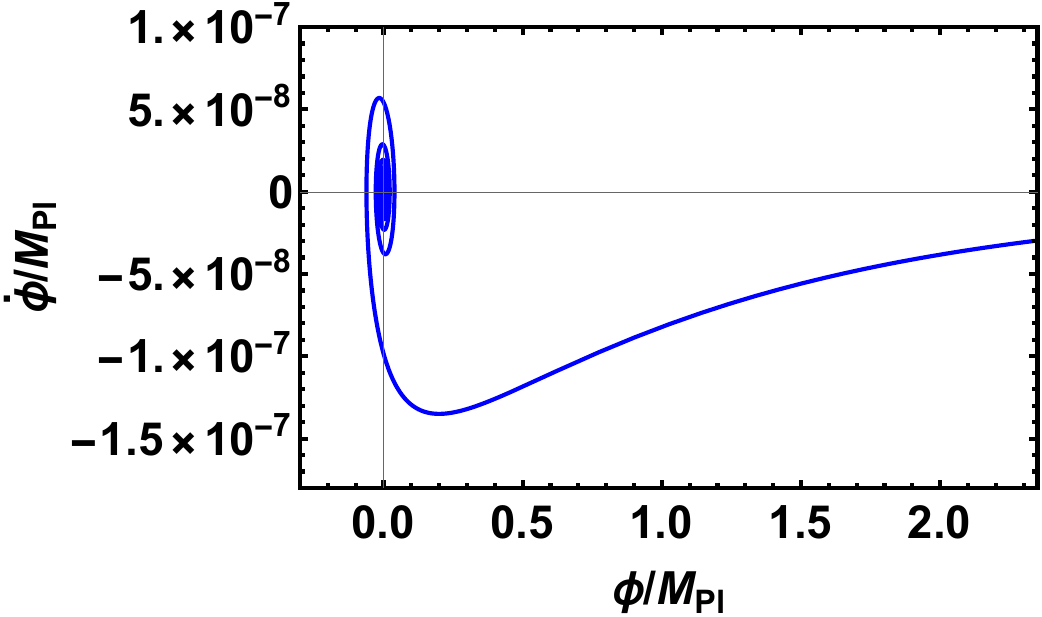}\\
    \vspace{2mm}
  \textbf{c)}
    \hspace{1mm}\includegraphics[width=.43\linewidth]{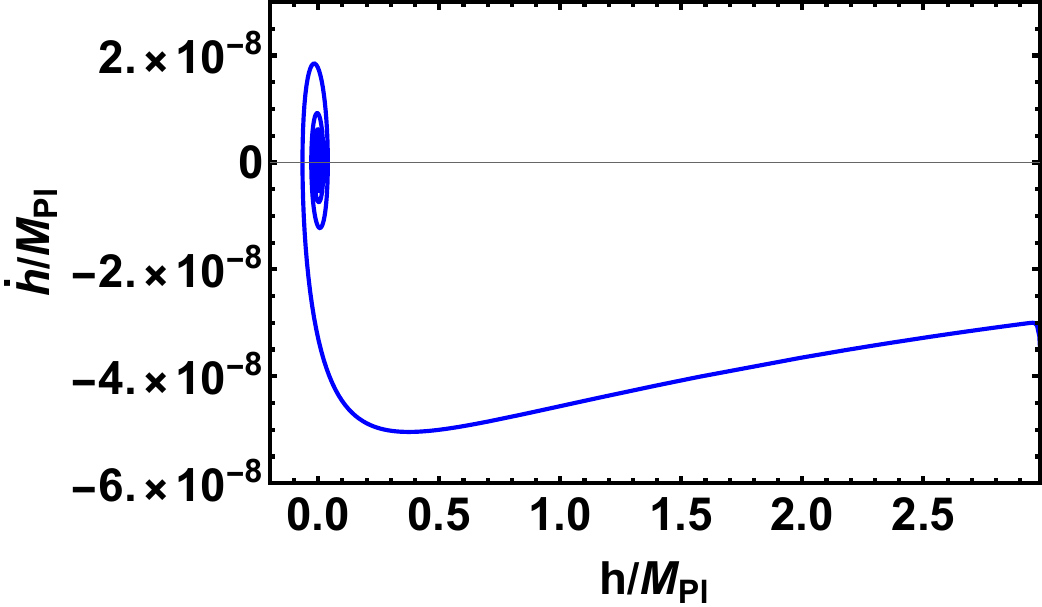}
  \hspace{0mm}
  \textbf{d)}
    \includegraphics[width=.43\linewidth]{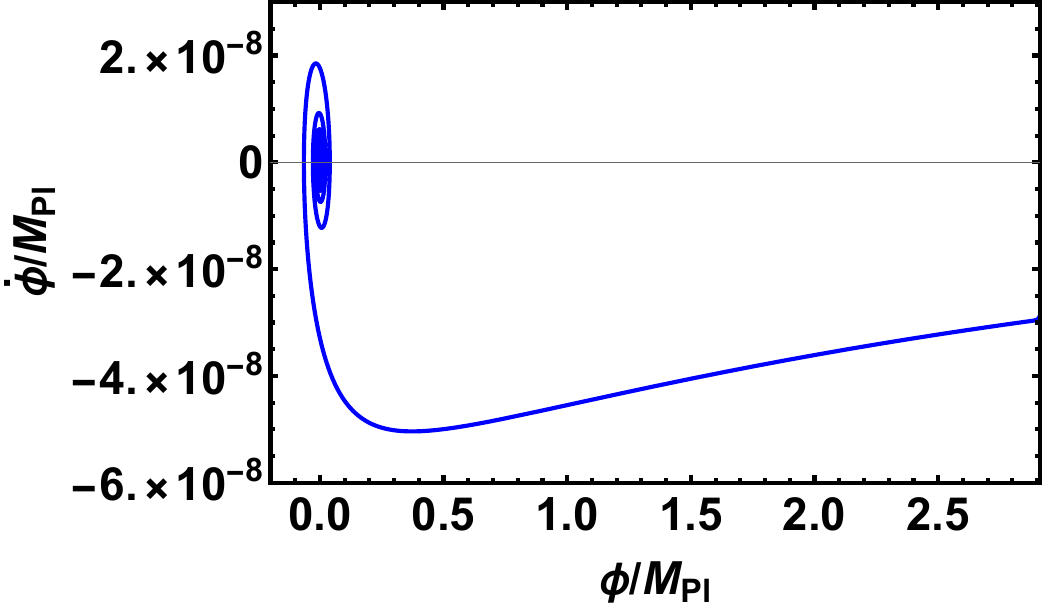}\\\vspace{2mm}
  \textbf{e)}\hspace{0mm}
    \includegraphics[width=.43\linewidth]{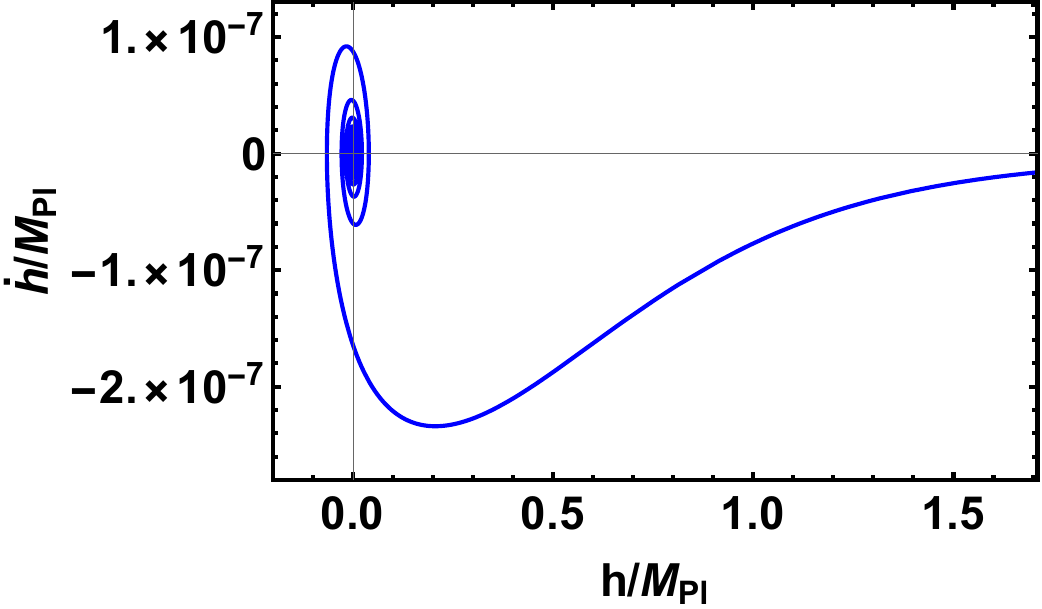}
  \hspace{0mm}
    \textbf{f)}\hspace{0mm}
    \includegraphics[width=.43\linewidth]{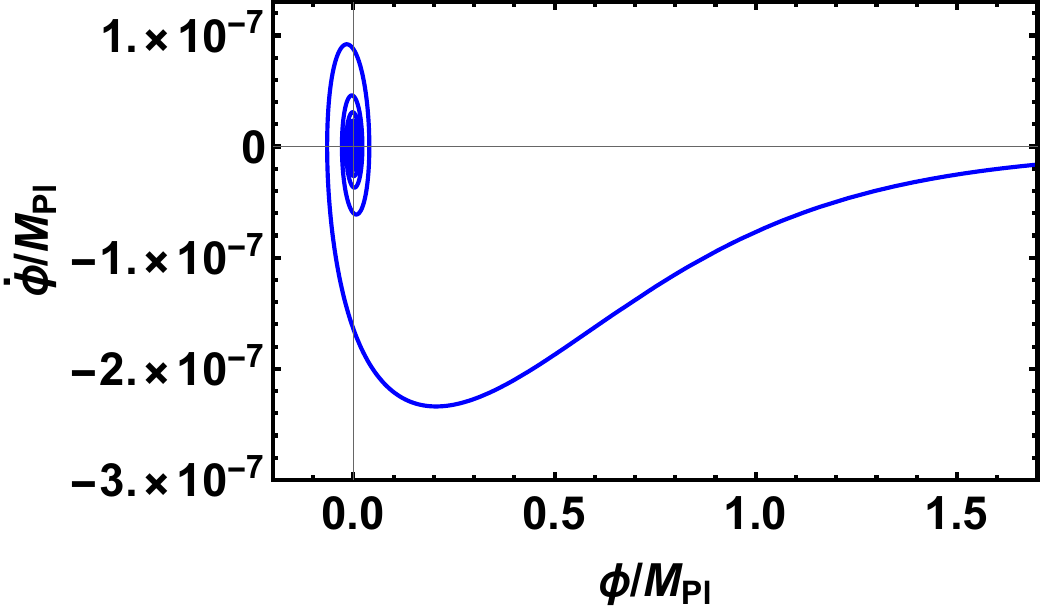}\\
    \vspace{2mm}
  \textbf{g)}\hspace{0mm}
    \includegraphics[width=.44\linewidth]{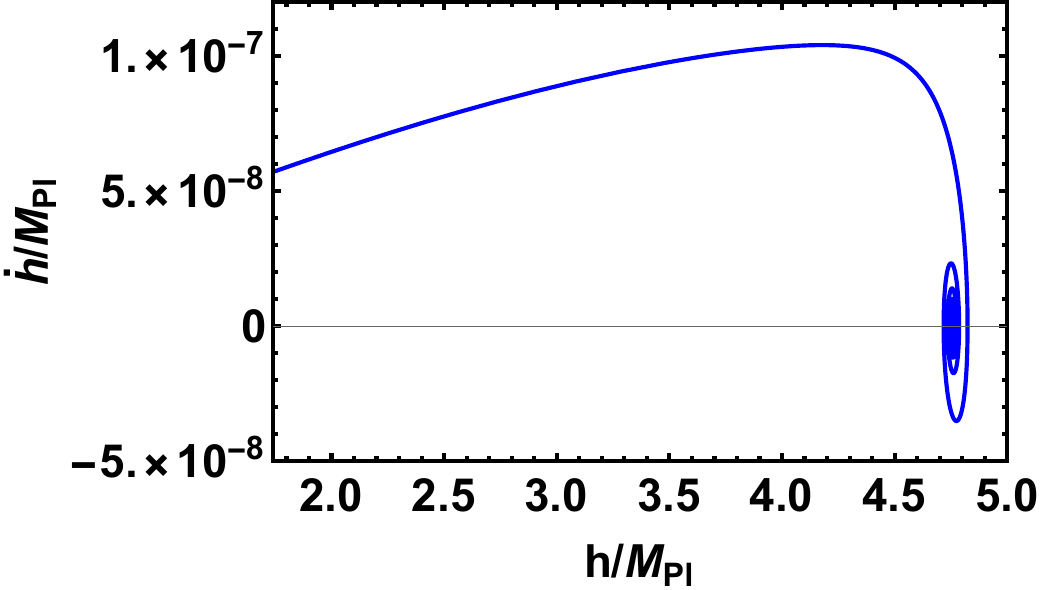}  \hspace{0mm}
    \textbf{h)}
    \includegraphics[width=.44\linewidth]{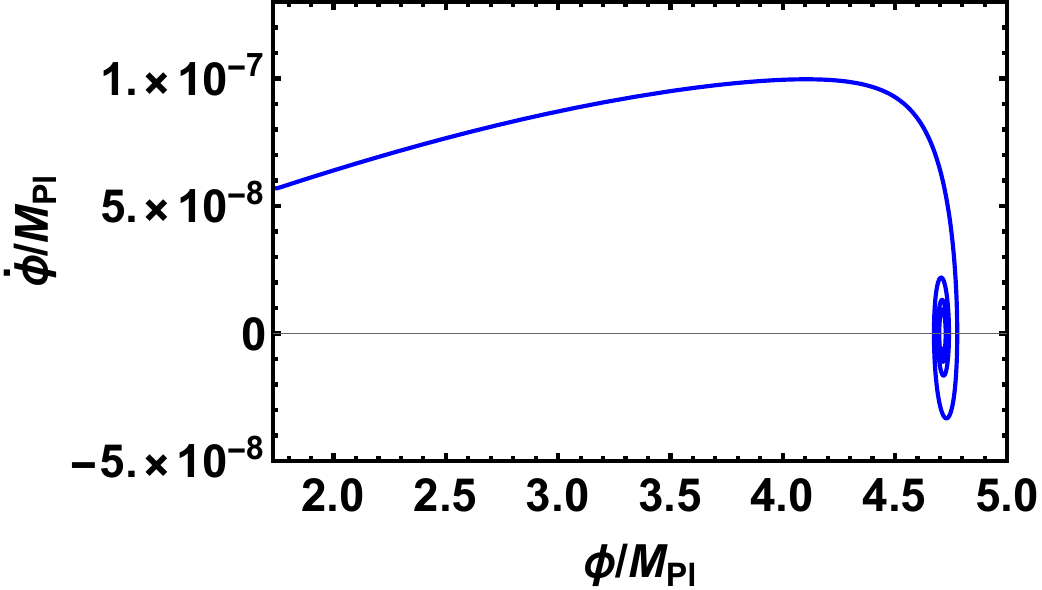}
  \caption{Inflation configuration space associated with the positive non-minimally coupled potentials introduced in Sect. \ref{Sect 2.3}: on the left in the Einstein frame, while on the right in the Jordan frame. \textbf{a), b)} the Starobinsky potential $V^E_S(h)$, from Eq. \eqref{star}; \textbf{c), d)} the $E-$model $V^E_E(h)$, from Eq. \eqref{E pot}, with $\alpha^E=10$; \textbf{e), f)} the $T-$model $V^E_T(h)$, from Eq. \eqref{T pot}, with $\alpha^T=0.1$; \textbf{g), h)} natural inflation potential $V^E_N(h)$, from Eq. \eqref{nat pot}, with $f=1.5M_{Pl}$.}
  \label{fig dyn}
\end{figure*}

Initially, we compare the potential patterns of the positive and negative coupling along with their respective slow-roll parameters in Fig. \ref{fig pot}. Then, we show the configuration spaces in both the Jordan and Einstein frames, Fig. \ref{fig dyn}. The attractors around $\phi_0$ and $h_0$ represent the inflationary graceful exit. To avoid redundancy, since we have proved that the inflationary features between negative and positive coupling are similar for small coupling constants, we focus on $\xi >0$. 

\begin{figure*}[p]
  \centering
  \textbf{a)}\hspace{0.mm}
    \includegraphics[width=.41\linewidth]{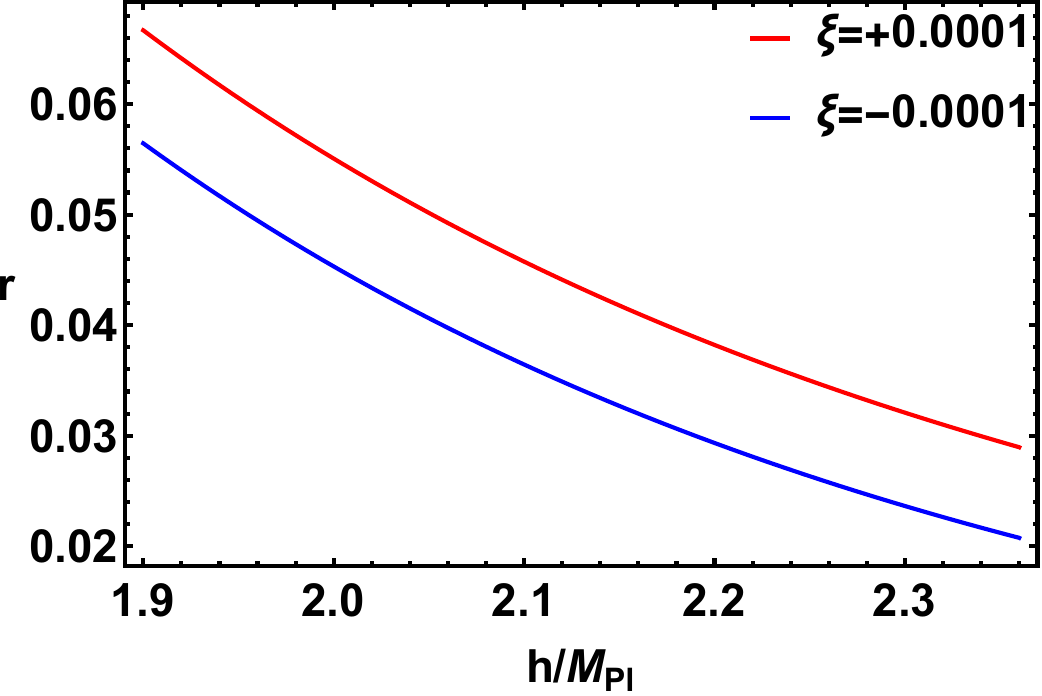}
  \hspace{0mm}
  \textbf{b)}\hspace{0mm}
    \includegraphics[width=.43\linewidth]{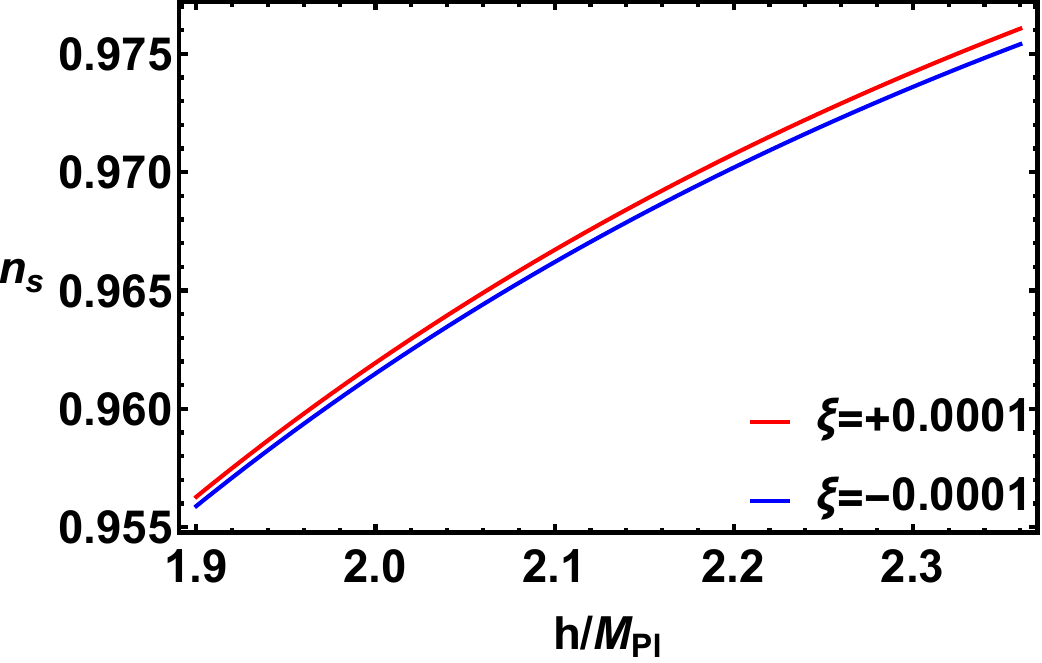}\\
    \vspace{0mm}
  \textbf{c)}
     \hspace{1mm}\includegraphics[width=.415\linewidth]{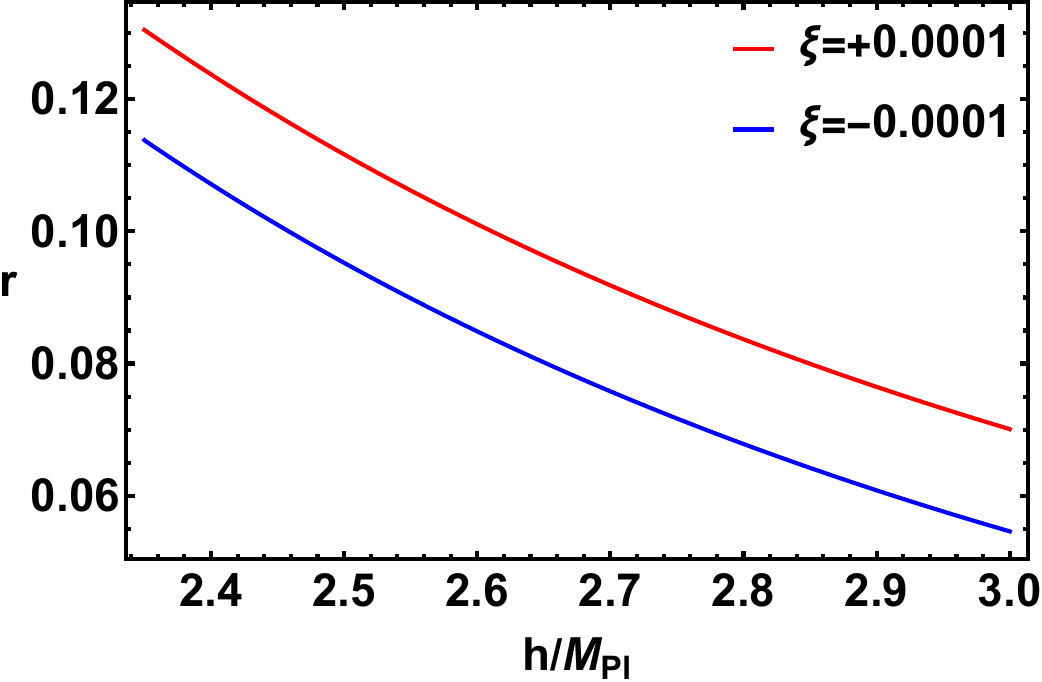}
  \hspace{0mm}
  \textbf{d)}
    \includegraphics[width=.43\linewidth]{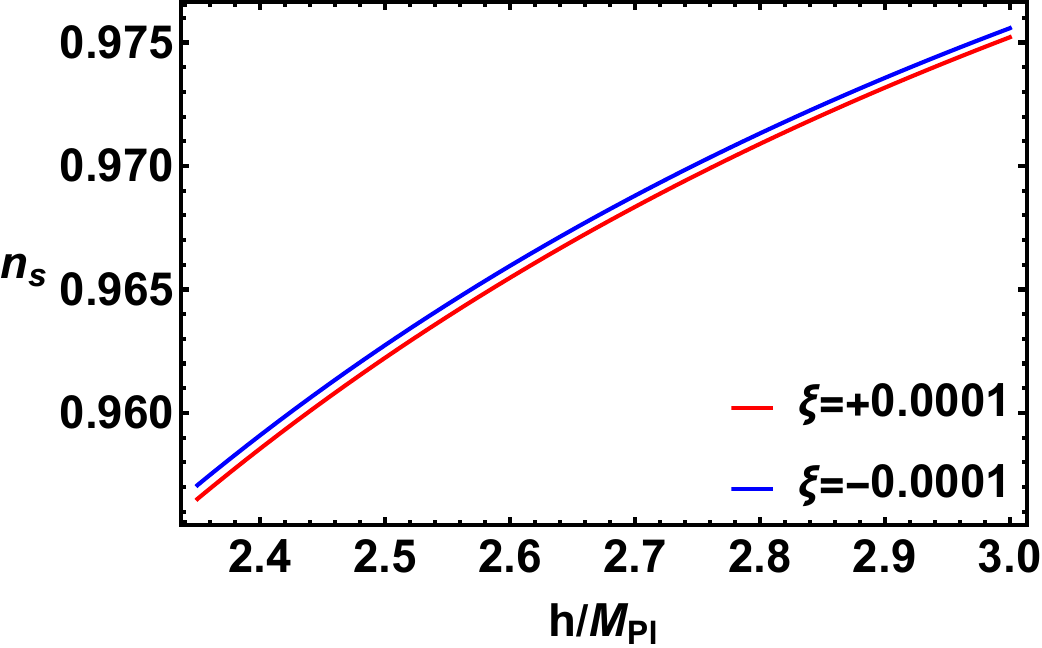}\\
  \textbf{e)}\hspace{0mm}
    \includegraphics[width=.42\linewidth]{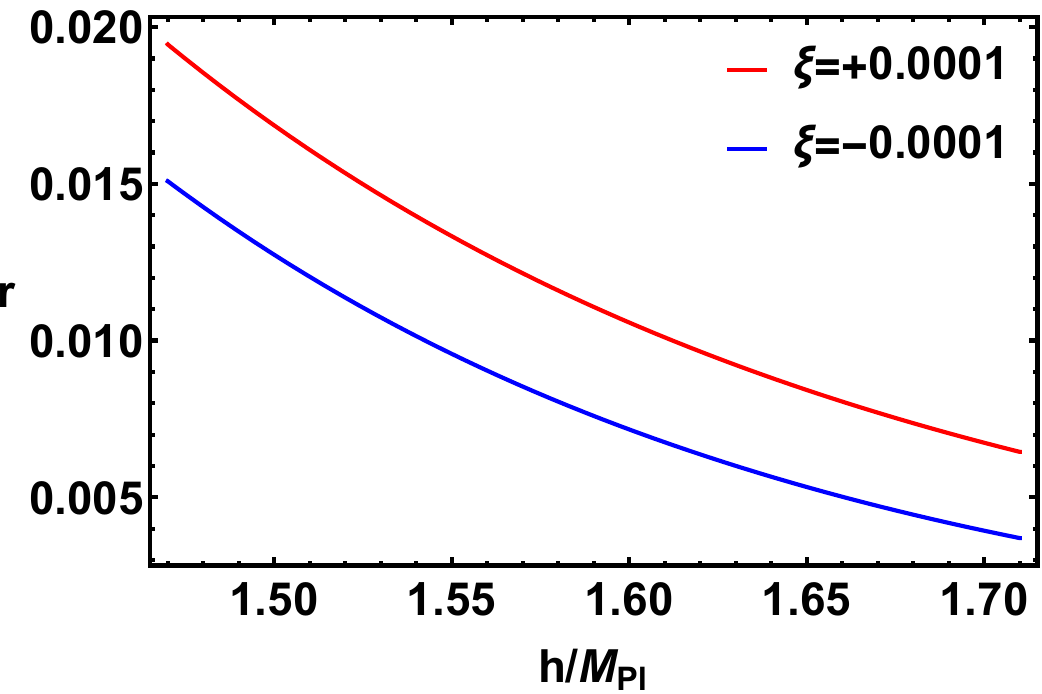}
  \hspace{0mm}
    \textbf{f)}\hspace{0mm}
    \includegraphics[width=.43\linewidth]{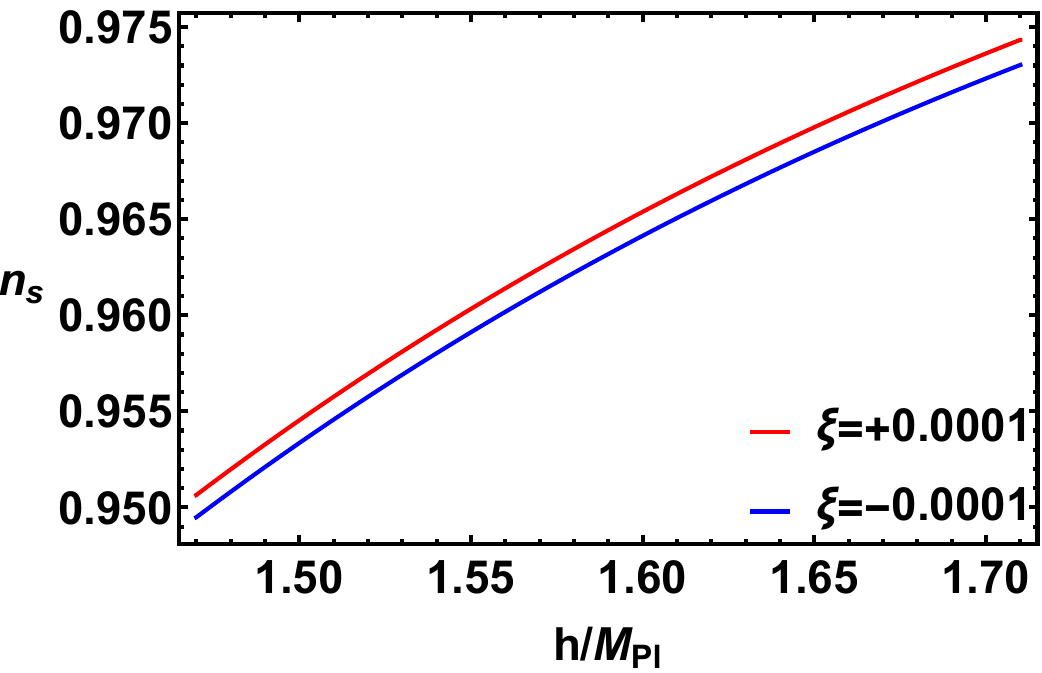}\\
    \vspace{0mm}
  \textbf{g)}\hspace{0mm}
    \includegraphics[width=.41\linewidth]{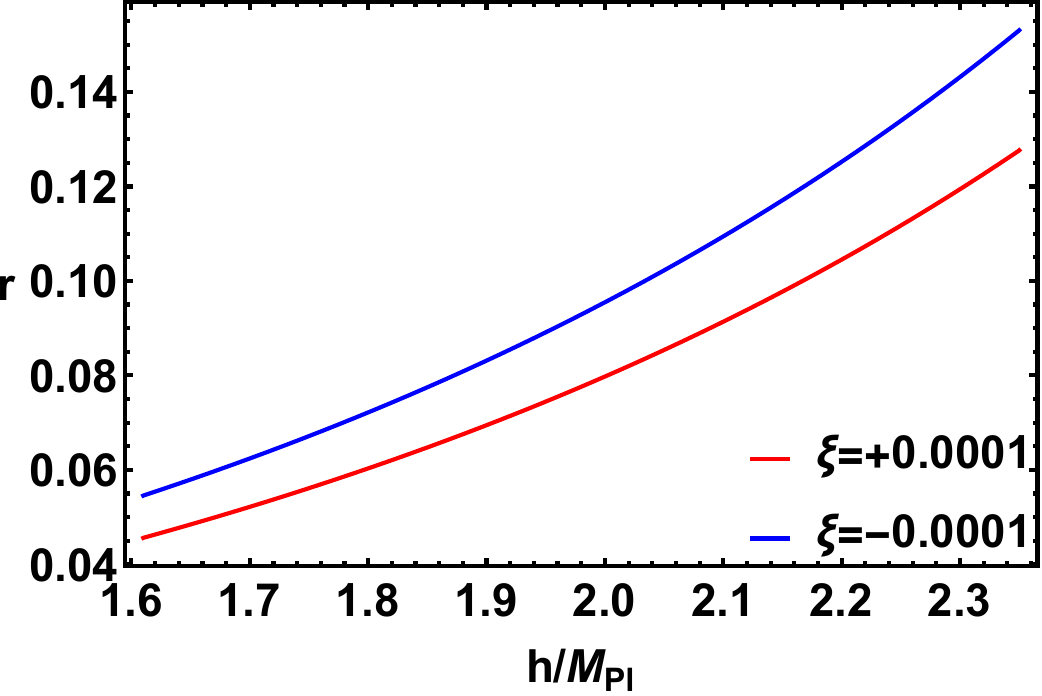}  \hspace{0mm}
    \textbf{h)}
    \includegraphics[width=.43\linewidth]{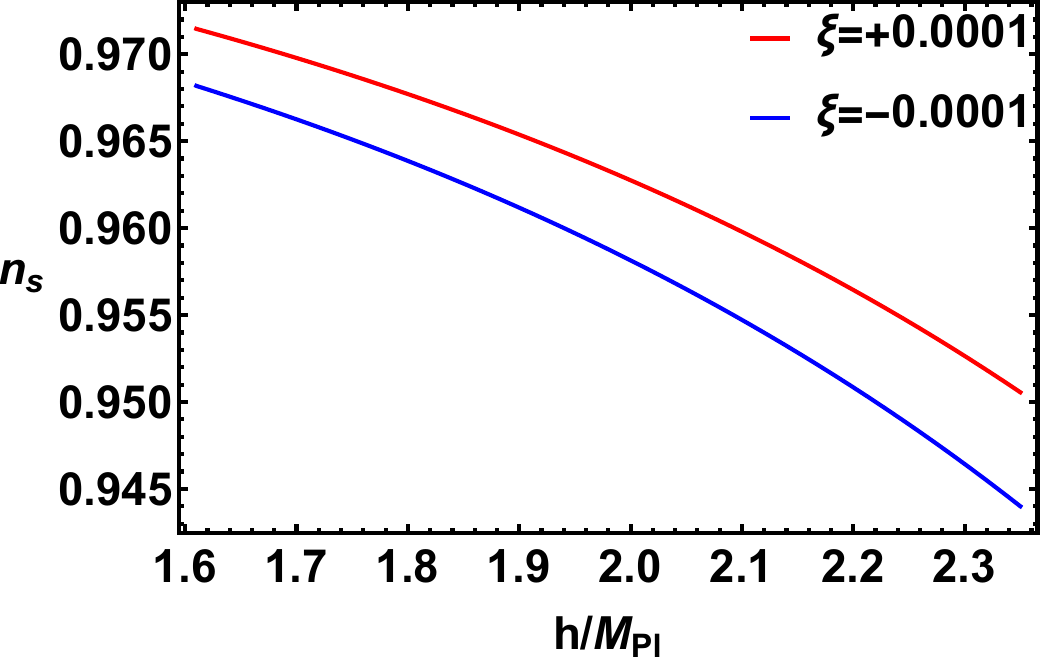}
  \caption{On the left the tensor-to-scalar ratio, $r$ defined in Eq. \eqref{r}, and on the right the spectral index, $n_s$ defined in Eq. \eqref{ns}, computed for positive and negative non-minimal coupling: \textbf{a), b)} the Starobinsky potential $V^E_S(h)$, from Eq. \eqref{star}; \textbf{c), d)} the $E-$model $V^E_E(h)$, from Eq. \eqref{E pot}, with $\alpha^E=10$; \textbf{e), f)} the $T-$model $V^E_T(h)$, from Eq. \eqref{T pot}, with $\alpha^T=0.1$; \textbf{g), h)} natural inflation potential $V^E_N(h)$, from Eq. \eqref{nat pot}, with $f=1.5M_{Pl}$.}
  \label{fig rns}
\end{figure*}

Finally, we compute the tensor-to-scalar ratio and the spectral index to ensure the viability of the non-minimally coupled extensions of the studied potentials with respect to the observational constraints. These parameters are functions of the e-folding number, ranging from $N\simeq70$, namely the onset of the slow-roll phase, to $N\simeq40$ In particular, Fig. \ref{fig rns} displays plots reporting the corresponding values of the field.

\end{document}